\documentclass[11pt]{article} 
\usepackage[left=2.5cm, right=2.5cm, top=1.785cm, bottom=2.0cm]{geometry}

\usepackage[utf8]{inputenc}

\usepackage{amsmath}
\usepackage{amsfonts}
\usepackage{amssymb}
\usepackage{amsthm}
\usepackage{mathtools}
\usepackage{mathrsfs}
\usepackage{nicefrac}

\usepackage[pdftex]{graphicx}
\usepackage{epstopdf}
\usepackage{subcaption}
\usepackage{float}
\usepackage{transparent}

\usepackage{booktabs}
\usepackage{tabularx}
\usepackage{longtable}
\usepackage{adjustbox}
\usepackage{siunitx}
\usepackage{multirow}
\usepackage{hhline}

\usepackage[tableposition=below]{caption}

\usepackage[
    pdfstartview=XYZ,
    bookmarks=true,
    colorlinks=true,
    linkcolor=blue,
    urlcolor=blue,
    citecolor=blue,
    pdftex,
    linktocpage=true,
    hyperindex=true
]{hyperref}
\usepackage{cleveref}
\crefname{equation}{Eq.}{Eqs.}

\usepackage{enumerate}

\usepackage{algorithm}
\usepackage{algorithmicx}
\usepackage[noend]{algpseudocode}

\usepackage{listings}
\usepackage{courier}

\usepackage{color}
\usepackage{setspace}
\usepackage{soul}
\usepackage{empheq}
\usepackage{cancel}
\usepackage{framed}
\usepackage{lineno}

\usepackage{lipsum}
\usepackage{import}
\usepackage[colorinlistoftodos]{todonotes}
\usepackage{authblk}
\usepackage{titlesec}
\usepackage[numbers,sort&compress]{natbib}

\theoremstyle{definition}

\definecolor{tbf}{RGB}{255,0,0}
\definecolor{txue}{RGB}{0,0,255}

\newcommand\keywords[1]{\textbf{Keywords}: #1}

\title{\Large Physics-guided diffusion models for inverse design of disordered metamaterials}

\begin{document}

\author[1]{\normalsize Ziyuan Xie}
\author[1]{\normalsize Weipeng Xu}
\author[1, 2]{\normalsize Dazhi Zhao}
\author[1]{\normalsize Wenchang Zhang}
\author[1]{\normalsize Daoyang Dong}
\author[1]{\normalsize Bingbing Xu}
\author[1]{\normalsize Ning Liu}
\author[3]{\normalsize Sheng Mao}
\author[1]{\normalsize Tianju Xue
\footnote{\textit{cetxue@ust.hk}  (corresponding author)}}

\affil[1]{\footnotesize Department of Civil and Environmental Engineering, The Hong Kong University of Science and Technology, Hong Kong, China}
\affil[2]{\footnotesize School of Aerospace Engineering and Applied Mechanics, Tongji University, Shanghai, China}
\affil[3]{\footnotesize Department of Mechanics, School of Mechanics and Engineering Science, Peking University, Beijing, China}

\date{}
\maketitle
\vspace{-20pt}

\begin{abstract}

Disordered metamaterials are promising for programming physical properties across diverse applications, yet their inverse design remains challenging due to the non-intuitive structure-property relationships and large design spaces. 
Recent generative approaches, particularly diffusion models, have shown potential in high-dimensional inverse design tasks. 
However, existing methods typically rely on carefully crafted training objectives, such as conditional data-driven or physics-informed loss functions. 
Because these strategies are inherently task-specific, the model must be retrained from scratch whenever the design problem changes (e.g., different governing equations, boundary conditions, or design objectives), severely limiting their flexibility and generalization ability.
In this work, we propose physics-guided diffusion models that leverage differentiable physics-based solvers to instantly guide the generative process for inverse design. 
Drawing inspiration from classifier guidance, we develop a sampling strategy that directly incorporates physics guidance into the reverse stochastic differential equations. 
Our approach enables task-adaptive generation using gradients from differentiable solvers, while the diffusion model itself needs to be trained only once on unlabeled data.
Focusing on disordered foam metamaterials, we present three representative design tasks: (1) achieving target effective thermal conductivity, (2) matching desired load–displacement response, and (3) maximizing energy absorption involving fractures. 
In each scenario, the proposed method successfully generates foam-like geometries that fulfill the prescribed physical objectives. 
These results demonstrate the versatility, efficiency, and practicality of physics-guided diffusion models for tackling complex inverse design problems in disordered metamaterials and beyond.

\end{abstract}



\section{Introduction}

Metamaterials have revolutionized the fields of mechanics, thermology, acoustics and optics by offering special material properties not found in nature~\cite{bertoldi2017flexible, hu2021thermal, cummer2016controlling, chen2010transformation, kadic20193d}. While periodic metamaterials have been extensively studied, disordered structures have recently attracted increasing attention due to their unique advantages, including enhanced defect and damage tolerance, greater robustness to fabrication variability, and increased multifunctionality and programmability, etc~\cite{denz2010nonlinearities, zaiser2023disordered}.
However, the inverse design of disordered metamaterials presents a formidable challenge. Unlike periodic structures where properties can be engineered through systematic modification of a few unit-cell parameters, disordered systems lack such clear design principles. The structure-property relationships are highly complex and non-intuitive, which makes designing disordered structures with targeted properties exceedingly difficult, rendering traditional trial-and-error approaches impractical.

Conventional inverse design methods for disordered metamaterials face considerable hurdles. Gradient-based approaches such as topology optimization (TO)~\cite{eschenauer2001topology} are susceptible to local optima, particularly in disordered systems~\cite{herrmann2024neural}. Additionally, gradient-based methods struggle to incorporate complex morphological priors, often producing free-form geometries that fail to match the required statistical or geometric characteristics. Gradient-free methods such as evolutionary algorithms~\cite{vikhar2016evolutionary} and simulated annealing~\cite{kirkpatrick1983optimization} offer better global exploration but become computationally prohibitive as the dimensionality of the design space grows.

Recent advances in generative models offer a paradigm shift for addressing these challenges in metamaterial inverse design. Generative models, including variational autoencoders (VAEs)~\cite{kingma2013auto}, generative adversarial networks (GANs)~\cite{goodfellow2014generative}, flow models~\cite{rezende2015variational} and diffusion models~\cite{ho2020denoising, song2020score}, learn the underlying distribution of training data and can synthesize new samples that preserve the statistical characteristics of the original dataset. Unlike conventional inverse design methods, these models naturally encode morphological priors through their training process, ensuring generated structures inherit realistic geometric features from existing designs. Moreover, generative models are inherently stochastic, capable of producing diverse candidate solutions and exploring the design space beyond local optima. In recent works, Shu et al.~\cite{shu2023physics} introduced a residual-guided diffusion model for flow-field reconstruction, and fidelity constraints were incorporated in a classifier-free manner~\cite{ho2022classifier}; Bastek et al.~\cite{bastek2024physics} proposed physics-informed diffusion models to solve partial differential equations and inverse design problems by embedding the physics residuals in the model training process; Ang et al.~\cite{ang2025deep} designed a framework based on GANs and physics solvers to design and evaluate synthetic disordered cellular structures; Baldan et al.~\cite{baldanphysics} developed a physics-based flow matching model which could solve inverse design problems with balanced data distribution and physics constraints. However, most of the current generative models for inverse design require labeled data or physical residuals in the training process, which often incurs substantial annotation effort or additional training cost. Most importantly, such models are typically restricted to a single type of design problem; adaptation to new design problems (e.g., different governing equations, boundary conditions, design objectives, etc.) generally requires retraining the entire model, thereby limiting their flexibility and practicality.

\begin{figure}[H]
    \centering
    \includegraphics[width=1.0\linewidth]{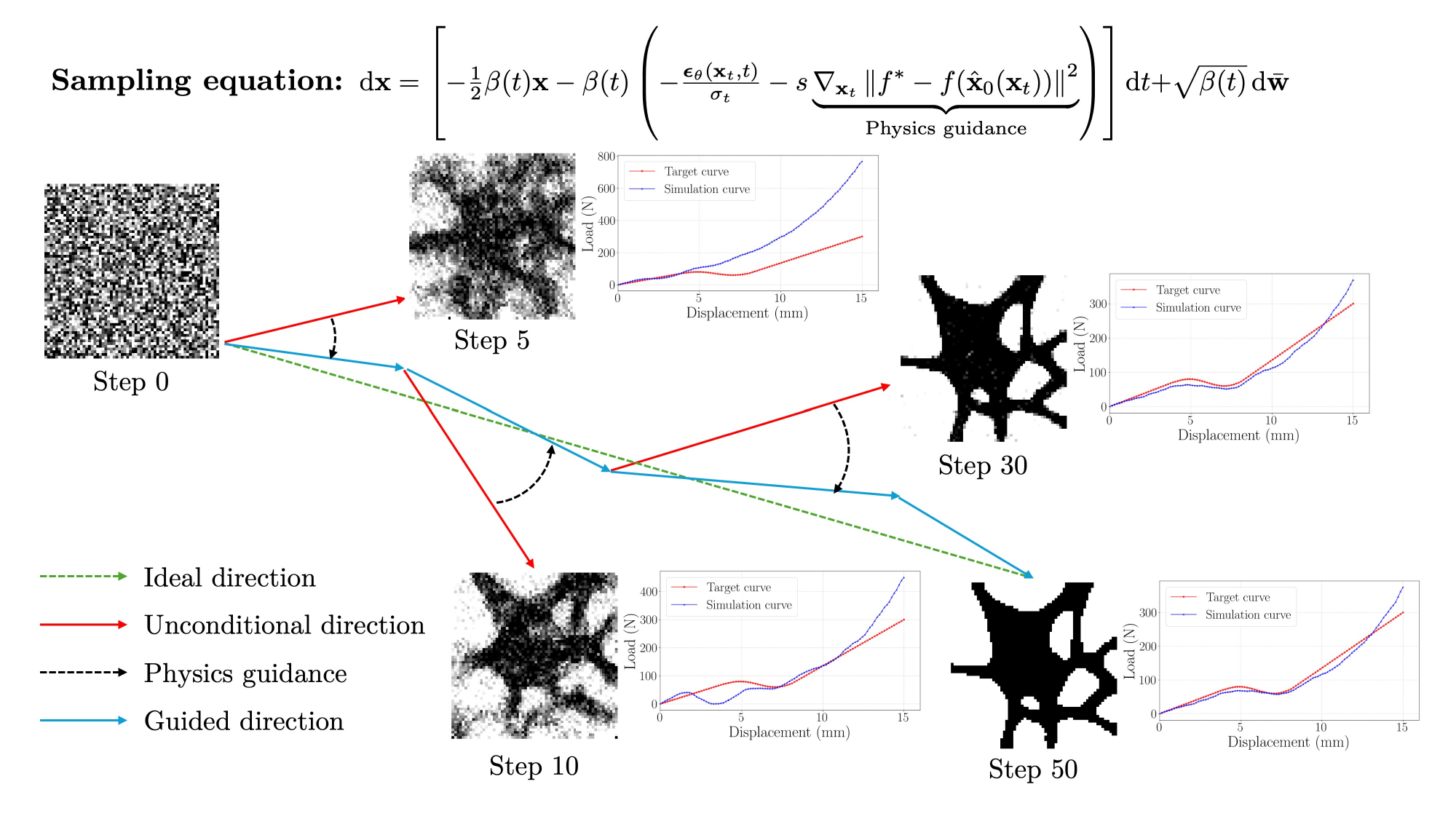}
    \caption{Illustration of physics-guided sampling process in diffusion models, where the goal is to achieve target load-displacement response.}
    \label{fig:intro}
\end{figure}

Classifier guidance has emerged as an effective strategy to steer diffusion models toward desired categories or attributes for conditional generation. However, training an additional classifier on noisy intermediate data is often difficult~\cite{dhariwal2021diffusion}. 
Herein, we propose physics-guided diffusion models that replace the classifier-based guidance term in the sampling process with physics-based guidance enabled by differentiable solvers~\cite{blondel2024elements, xue2023jax}. 
For demonstration, an inverse design process with physics guidance leading to a structure with prescribed load-displacement response is shown in Fig.~\ref{fig:intro}. 
This framework brings three key advantages. First, the diffusion model can be trained on unlabeled datasets, substantially reducing the effort required for data preparation. Second, the optimization is driven by physics-based objectives rather than learned approximations, improving accuracy and reliability. Third, a well-trained model can be readily adapted to new design specifications or multiple objectives without retraining the model. We validate the effectiveness and flexibility of the proposed method through three diverse numerical case studies on inverse design of disordered metamaterials, including (1) achieving target effective thermal conductivity, (2) matching desired load–displacement response, and (3) maximizing energy absorption involving fractures under a prescribed volume-fraction constraint. 
In this work, we focus on the inverse design of disordered closed-cell foam materials, but the methodology can be applied to other engineering design problems in a straightforward manner.

The rest of this paper is organized as follows: Section~\ref{sec:methods} describes details of the proposed physics-guided diffusion model. Section~\ref{sec:case} shows three numerical cases to demonstrate the efficacy and adaptability of this inverse design method. Section~\ref{sec:disc} presents a discussion of this work. Finally, Section~\ref{sec:conclusion} summarizes the contents and contributions of this study.

\section{Physics-guided diffusion models}

In this section, we first describe the procedure of generating disordered closed-cell foam images via Voronoi tessellation. We then review classical score-based generative models formulated with SDEs, as well as classifier-guidance diffusion models. Then, we present our physics-guided diffusion models for inverse design.

\label{sec:methods}

\subsection{Voronoi tessellation}

In this study, the objective is to design disordered closed-cell foam materials. To construct the dataset of disordered foams, we employ the Voronoi tessellation method~\cite{du1999centroidal}, which partitions a plane into regions based on distance to a specific set of seed points. Voronoi tessellation naturally generates cellular structures, making it an ideal choice for modeling disordered foam structures~\cite{wang2023note}.

\begin{figure}[H]
    \centering
    \includegraphics[width=1\linewidth]{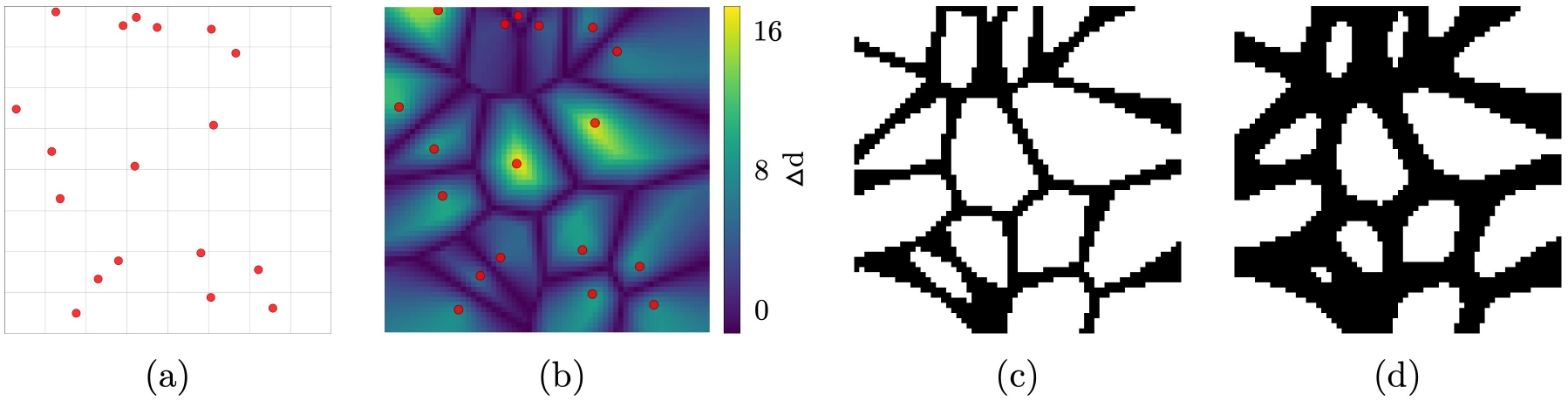}
    \caption{Steps for generating a disordered closed-cell foam structure: (a) Randomly distribute seed points; (b) Compute the distance field; (c) Threshold the distance field to obtain the cell skeleton; (d) Adjust the wall thickness.}
    \label{fig:von}
\end{figure}

We generate foam structures using an implicit distance-field formulation of Voronoi tessellation~\cite{borgefors1986distance}. The process begins by randomly distributing $N = (H / d_{\text{cell}})^2$ seed points in an $H \times W$ domain, where $d_{\text{cell}}$ is a prescribed cell diameter, and $H = W$ in this work. For each pixel, we compute the distances $d_1$ and $d_2$ to its two nearest seeds via the k-dimensional tree algorithm~\cite{bentley1975multidimensional}. The difference $\Delta d = d_2 - d_1$ acts as a Voronoi boundary detector, and pixels with $\Delta d < t_{\text{init}}$ form the initial wall skeleton, where $t_{\text{init}}$ is the initial wall thickness. This thresholding operation implicitly extracts Voronoi cell boundaries as finite-width walls rather than infinitesimal geometric lines, naturally accommodating material thickness without explicit polygon offsetting. To transform the sharp, pixelated skeleton into smooth, organic foam morphology, we apply Gaussian convolution with standard deviation $\sigma = \alpha \cdot d_{\text{cell}}$, where $\alpha$ is the roundness factor which is set to 0.1 in this work. By linking $\sigma$ to $d_{\text{cell}}$, we can ensure that corner roundness scales proportionally with cell size, maintaining consistent visual appearance across different resolutions. Finally, we binarize the smoothed field to achieve the target volume fraction $\phi$. The steps of the foam generation are illustrated in Fig.~\ref{fig:von}. We vary $d_{\text{cell}}$ (from 6 to 20 pixels) and $\phi$ (from 0.3 to 0.8) to generate 6,400 disordered closed-cell foam binary images with a resolution of $64 \times 64$ for training.

\subsection{Physics-guided diffusion models}

\subsubsection{Score-based generative models} 
\label{Sec:sde}

Song et al.~\cite{song2020score} unified two successful classes of probabilistic generative models, score matching with Langevin dynamics (SMLD)~\cite{song2020sliced} and denoising diffusion probabilistic modeling (DDPM)~\cite{ho2020denoising}, by using score-based generative framework through stochastic differential equations (SDEs). We adopt this score-based generative framework in this work, as the SDE formulation enables flexible and controllable data generation.

Following Song et al.~\cite{song2020score}, we consider a forward process that perturbs original data $\mathbf{x}(0) \sim p_0$ into $\mathbf{x}(T) \sim p_T$ over a continuous time interval $t \in [0, T]$, where $p_T$ is a Gaussian distribution with fixed mean and variance. This forward diffusion process is described by the following SDE:
\begin{equation}
    \mathrm{d}\mathbf{x} = \mathbf{f}(\mathbf{x}, t) \mathrm{d}t + g(t) \mathrm{d}\mathbf{w}
\end{equation}
where $\mathbf{f}(\cdot, t): \mathbb{R}^d \to \mathbb{R}^d$ is a vector-valued function called the drift coefficient of $\mathbf{x}(t)$, $g(\cdot): \mathbb{R} \to \mathbb{R}$ is a scalar function called the diffusion coefficient of $\mathbf{x}(t)$, and $\mathbf{w}$ is a standard Wiener process. In this work, we employ the variance-preserving (VP) SDE:
\begin{equation}
    \mathrm{d}\mathbf{x} = -\frac{1}{2}\beta(t) \mathbf{x} \mathrm{d}t + \sqrt{\beta(t)} \mathrm{d}\mathbf{w}
\end{equation}
where $\beta(t)$ is a monotonically increasing noise schedule function. The VP-SDE has the key property that the marginal distribution $p_t(\mathbf{x}_t | \mathbf{x}_0)$ is analytically tractable:
\begin{equation}
    p_t(\mathbf{x}_t | \mathbf{x}_0) = \mathcal{N}\left(\mathbf{x}_t; \sqrt{\bar{\alpha}_t} \mathbf{x}_0, (1 - \bar{\alpha}_t) \mathbf{I}\right)
\end{equation}
where $\bar{\alpha}_t = \exp\left(-\int_0^t \beta(s) \mathrm{d}s\right)$.

\begin{figure}[H]
    \centering
    \includegraphics[width=0.9\linewidth]{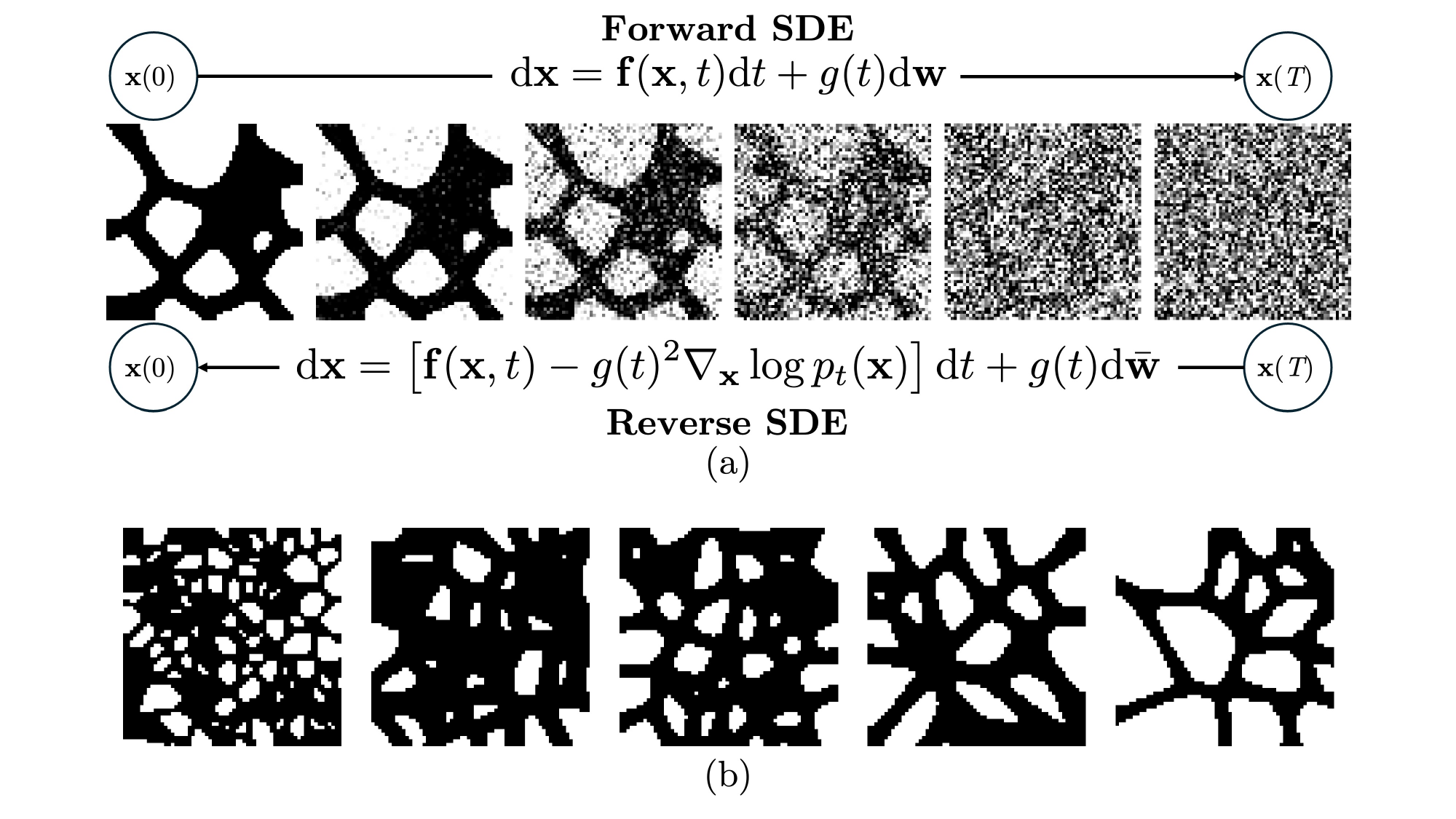}
    \caption{(a) Illustration of the forward and reverse processes described by SDEs; (b) Examples of generated foams.}
    \label{fig:sde}
\end{figure}

The reverse process is obtaining the samples $\mathbf{x}(0) \sim p_0$ by starting from prior data distribution $\mathbf{x}(T) \sim p_T$, which is also a diffusion process~\cite{anderson1982reverse} and can be given by the reverse SDE:
\begin{equation}
    \mathrm{d}\mathbf{x} = \left[\mathbf{f}(\mathbf{x}, t) - g(t)^2 \nabla_{\mathbf{x}} \log p_t(\mathbf{x})\right] \mathrm{d}t + g(t) \mathrm{d}\bar{\mathbf{w}}
\end{equation}
where $\mathrm{d}t$ now flows backward from $T$ to $0$, $\mathrm{d}\bar{\mathbf{w}}$ is a reverse-time Wiener process, and $\nabla_{\mathbf{x}} \log p_t(\mathbf{x})$ is the score function of the marginal density $p_t(\mathbf{x})$. Fig.~\ref{fig:sde}~(a) is an illustration of the score-based generative workflow through SDEs.

The score function $\nabla_{\mathbf{x}} \log p_t(\mathbf{x})$ is approximated by a time-dependent neural network $\mathbf{s}_\theta(\mathbf{x}, t)$. Given the transition distribution $p_t(\mathbf{x}_t | \mathbf{x}_0) = \mathcal{N}(\mathbf{x}_t; \sqrt{\bar{\alpha}_t} \mathbf{x}_0, \sigma_t^2 \mathbf{I})$ where $\sigma_t^2 = 1 - \bar{\alpha}_t$, we can write $\mathbf{x}_t = \sqrt{\bar{\alpha}_t} \mathbf{x}_0 + \sigma_t \boldsymbol{\epsilon}$ with $\boldsymbol{\epsilon} \sim \mathcal{N}(\mathbf{0}, \mathbf{I})$. The score function admits the following representation:
\begin{equation}
    \nabla_{\mathbf{x}_t} \log p_t(\mathbf{x}_t) = -\frac{\mathbb{E}[\boldsymbol{\epsilon} | \mathbf{x}_t]}{\sigma_t} = -\frac{\boldsymbol{\epsilon}_\theta(\mathbf{x}_t, t)}{\sigma_t}
\end{equation}
where $\boldsymbol{\epsilon}_\theta(\mathbf{x}_t, t)$ is a neural network trained to predict the noise $\boldsymbol{\epsilon}$ from $\mathbf{x}_t$. This equivalence between score matching and noise prediction enables efficient training via denoising score matching~\cite{ho2020denoising, song2020score}.

Substituting the VP-SDE drift coefficient and the noise-based score representation into the reverse SDE, we can obtain the sampling equation:
\begin{equation}
    \mathrm{d}\mathbf{x} = \left[-\frac{1}{2}\beta(t) \mathbf{x} + \beta(t) \frac{\boldsymbol{\epsilon}_\theta(\mathbf{x}_t, t)}{\sigma_t}\right] \mathrm{d}t + \sqrt{\beta(t)} \mathrm{d}\bar{\mathbf{w}}
\label{Eq:uncond-sde}
\end{equation}
This SDE can be solved numerically using standard SDE solvers from $t=T$ to $t=0$ to generate samples from the learned data distribution. 

A U-Net~\cite{ronneberger2015u} architecture is adopted to predict the noise $\boldsymbol{\epsilon}_\theta(\mathbf{x}_t, t)$. The U-Net follows a symmetrical encoder-decoder architecture consisting of four downsampling and four upsampling blocks. Feature channels expand progressively as (64, 128, 256, 512) to encode multi-scale representations. Each block incorporates two residual blocks for adequate feature learning. Self-attention mechanisms are selectively applied at the third and fourth blocks (256 and 512 channels) to balance efficiency and representation power, utilizing 4 attention heads with 32-group normalization. Temporal information is encoded through sinusoidal embeddings and subsequently transformed via a two-layer multilayer perceptron into a 256-dimensional conditioning vector. This time encoding is additively integrated into every residual block, enabling global temporal guidance throughout the network hierarchy. The prepared 6,400 binary foam images are used to train this U-Net, and some generated samples are shown in Fig.~\ref{fig:sde}~(b).

\subsubsection{Classifier guidance}

The score-based model described above is unconditional, so it can only generate random samples from the learned data distribution $p(\mathbf{x})$. However, for conditional generation tasks, we seek to generate data that have specific attributes. Classifier guidance~\cite{dhariwal2021diffusion} provides a principled approach to guide the generative process to desired categories.

The core idea of classifier guidance is to leverage a pre-trained classifier $p_\kappa(y | \mathbf{x})$ to steer the reverse diffusion process toward a target class $y^*$. By Bayes' rule, the conditional score $\nabla_{\mathbf{x}_t} \log p(\mathbf{x}_t | y^*)$ can be decomposed as:
\begin{equation}
    \nabla_{\mathbf{x}_t} \log p(\mathbf{x}_t | y^*) = \nabla_{\mathbf{x}_t} \log p(\mathbf{x}_t) + \nabla_{\mathbf{x}_t} \log p(y^* | \mathbf{x}_t)
\end{equation}
where the first term on the right-hand side of this equation is the unconditional score which has been introduced in Section~\ref{Sec:sde}, while the second term $\nabla_{\mathbf{x}_t} \log p(y^* | \mathbf{x}_t)$ represents the gradient of the log-probability that the classifier assigns to label $y^*$ given the input $\mathbf{x}_t$, with respect to $\mathbf{x}_t$.

To apply classifier guidance, the classifier must be trained on noisy samples at various noise levels, so it learns to predict categories from partially perturbed data $\mathbf{x}_t$. The conditional reverse SDE then becomes:
\begin{equation}
    \mathrm{d}\mathbf{x} = \left[\mathbf{f}(\mathbf{x}, t) - g(t)^2 \left(\nabla_{\mathbf{x}_t} \log p_t(\mathbf{x}_t) + s \nabla_{\mathbf{x}_t} \log p_\kappa(y^* | \mathbf{x}_t)\right)\right] \mathrm{d}t + g(t) \mathrm{d}\bar{\mathbf{w}}
\label{Eq:class-sde}
\end{equation}
where $s$ is a guidance scale parameter that controls how strongly the classifier influences the generation process.

While classifier guidance has proven effective for categorical generation tasks in computer vision~\cite{dhariwal2021diffusion}, it faces challenges when applied to material inverse design problems. First, training accurate classifiers on noisy intermediate structures is difficult. Second, the classifier introduces approximation errors that can accumulate throughout the iterative generative process, potentially leading to physically inconsistent structures. Third, modifying design specifications or adding new optimization objectives requires retraining the classifier on appropriately labeled data, which is both computationally expensive and labor-intensive. Fourth, many inverse design problems are inherently regression tasks rather than classification problems. 

These limitations motivate the development of our physics-guidance sampling strategy in diffusion models, where the classifier is replaced with direct physics-based simulations to provide more accurate and reliable guidance signals.

\subsubsection{Physics guidance}

To overcome the limitations of classifier guidance, we propose physics-guided diffusion models that directly incorporate physics-based simulations into the generative process. Our objective is to generate disordered closed-cell foam structures $\mathbf{x}$ that achieve specific target physical properties $f^*$.

Following the conditional sampling framework, we begin with Bayes' rule. The score of the physics-constrained distribution can be decomposed as:
\begin{equation}
    \nabla_{\mathbf{x}_t} \log p(\mathbf{x}_t | f^*) = \nabla_{\mathbf{x}_t} \log p(\mathbf{x}_t) + \nabla_{\mathbf{x}_t} \log p(f^* | \mathbf{x}_t)
\label{Eq:bayes_phy}
\end{equation}
where the second term on the right-hand side of the equation represents the gradient of log probability $\log p(f^* | \mathbf{x}_t)$ with respect to the input $\mathbf{x}_t$. Unlike classifier guidance which relies on a learned surrogate model, we aim to construct this second term directly from physics simulations.

The ultimate goal in this work is to obtain a clean, noise-free image that can be fed into the physics solver $f(\cdot)$ to produce accurate results. However, during the reverse diffusion process we only have access to noisy intermediate states $\mathbf{x}_t$, which can introduce numerical errors and may even cause the solver to fail to converge. To directly optimize the final state $\mathbf{x}_0$, we use the estimation of clean data from the noisy observation to perform the physics-based calculations, which is given as~\cite{efron2011tweedie}:
\begin{equation}
    \hat{\mathbf{x}}_0 = \mathbb{E}[\mathbf{x}_0 | \mathbf{x}_t] = \frac{\mathbf{x}_t - \sigma_t \boldsymbol{\epsilon}_\theta(\mathbf{x}_t, t)}{\sqrt{\bar{\alpha}_t}}
\end{equation}

This predicted $\hat{\mathbf{x}}_0$ serves as a differentiable proxy for the clean structure, enabling us to evaluate physics-based objectives during the noisy sampling process. By the law of total probability, we have: 
\begin{equation}
p(f^* | \mathbf{x}_t) = \int p(f^* | \mathbf{x}_0)\, p(\mathbf{x}_0 | \mathbf{x}_t) \, d\mathbf{x}_0
\label{eq:total_p}
\end{equation}
where $p(\mathbf{x}_0 | \mathbf{x}_t)$ can be approximated by a Dirac delta function $\delta(\mathbf{x}_0 - \hat{\mathbf{x}}_0)$, as $\hat{\mathbf{x}}_0$ is the estimated mean of $p(\mathbf{x}_0 | \mathbf{x}_t)$. Then Eq.~(\ref{eq:total_p}) can be further written as: 
\begin{equation}
p(f^* | \mathbf{x}_t) \approx \int p(f^* | \mathbf{x}_0)\, \delta(\mathbf{x}_0 - \hat{\mathbf{x}}_0) \, d\mathbf{x}_0
= p(f^* | \hat{\mathbf{x}}_0)
\end{equation}

The approximation $p(f^* | \mathbf{x}_t) \approx p(f^* | \hat{\mathbf{x}}_0)$ becomes increasingly accurate as the posterior $p(\mathbf{x}_0|\mathbf{x}_t)$ concentrates during later stages of the reverse process (small $t$). While the approximation is less precise at early stages (large $t$), it empirically provides effective guidance throughout sampling~\cite{chung2022diffusion}. Considering a measurement model $f^* = f(\hat{\mathbf{x}}_0) + n$ with $n \sim \mathcal{N}(0, \tau)$, we have:
\begin{equation}
p(f^* | \hat{\mathbf{x}}_0) =
\dfrac{1}{\sqrt{2 \pi \tau}}
\exp\left[
    -\dfrac{ \left\lVert f^* - f(\hat{\mathbf{x}}_0) \right\rVert^2 }
    {2\tau}
\right]
\end{equation}

Then the second term on the right-hand side of Eq.~(\ref{Eq:bayes_phy}) can be written as:
\begin{equation}
    \nabla_{\mathbf{x}_t} \log p(f^* | \mathbf{x}_t) = -\frac{1}{2 \tau} \nabla_{\mathbf{x}_t} \|f^* - f(\hat{\mathbf{x}}_0(\mathbf{x}_t))\|^2
\end{equation}

Combining with the unconditional and classifier-guided reverse SDEs (Eq.~(\ref{Eq:uncond-sde}) and (\ref{Eq:class-sde})), we obtain the physics-guided sampling SDE for the VP diffusion model:
\begin{equation}
    \mathrm{d}\mathbf{x} = \left[
        -\frac{1}{2}\beta(t) \mathbf{x} 
        - \beta(t) \left(
            -\frac{\boldsymbol{\epsilon}_\theta(\mathbf{x}_t, t)}{\sigma_t} 
            - s \underbrace{\nabla_{\mathbf{x}_t} \left\|f^* - f(\hat{\mathbf{x}}_0(\mathbf{x}_t)) \right\|^2}_\text{Physics guidance}
        \right)
    \right] \mathrm{d}t 
    + \sqrt{\beta(t)}\, \mathrm{d}\bar{\mathbf{w}}
\label{eq:coreeq}
\end{equation}
where $s = 1 / 2 \tau$, which can be regarded as a coefficient controlling the intensity of guidance. In practice, we normalize the guidance term to ensure numerical stability during sampling. 

Alternatively, we can apply the chain rule to transform the gradient from $\mathbf{x}_t$ space to $\hat{\mathbf{x}}_0$ space:
\begin{equation}
    \nabla_{\mathbf{x}_t} \log p(f^* | \mathbf{x}_t) = - s \frac{\partial \hat{\mathbf{x}}_0(\mathbf{x}_t)}{\partial \mathbf{x}_t} \nabla_{\hat{\mathbf{x}}_0} \|f^* - f(\hat{\mathbf{x}}_0(\mathbf{x}_t))\|^2
\end{equation}

Then the  physics-guided sampling equation can be expressed as:
\begin{equation}
    \mathrm{d}\mathbf{x} = \left[
        -\frac{1}{2}\beta(t) \mathbf{x} 
        - \beta(t) \left(
            -\frac{\boldsymbol{\epsilon}_\theta(\mathbf{x}_t, t)}{\sigma_t} 
            - s \frac{\partial \hat{\mathbf{x}}_0}{\partial \mathbf{x}_t} \nabla_{\hat{\mathbf{x}}_0} \left\|f^* - f(\hat{\mathbf{x}}_0)\right\|^2
        \right)
    \right] \mathrm{d}t 
    + \sqrt{\beta(t)}\, \mathrm{d}\bar{\mathbf{w}}
\end{equation}
where the Jacobian $\partial \hat{\mathbf{x}}_0 / \partial \mathbf{x}_t$ is given as:
\begin{equation}
    \frac{\partial \hat{\mathbf{x}}_0}{\partial \mathbf{x}_t} = \frac{1}{\sqrt{\bar{\alpha}_t}}\left(\mathbf{I} - \sigma_t \frac{\partial \boldsymbol{\epsilon}_\theta(\mathbf{x}_t, t)}{\partial \mathbf{x}_t}\right)
\end{equation}

The U-Net training, physics-based simulations and gradient computations in this study are all implemented with \texttt{JAX}~\cite{jax2018github}, a high-performance numerical computing library with built-in GPU acceleration and automatic differentiation (AD) capabilities, to ensure seamless integration of physics solvers with the diffusion model and maintain computational efficiency throughout the physics-guided sampling process.
In particular, the gradient $\nabla_{\mathbf{x}_t} \left\|f^* - f(\hat{\mathbf{x}}_0(\mathbf{x}_t)) \right\|^2$ is provided directly by differentiable physics simulators. 
As higher-fidelity physical models become available, our framework can leverage them to tackle increasingly complex physics, without being tied to any specific solver.
In the following section, we demonstrate this advantage through a set of numerical case studies using differentiable finite element method (FEM) or differentiable material point method (MPM).

\subsubsection{Projection and model parameters}

The estimation $\hat{\mathbf{x}}_0$ is often noisy, particularly in the early iterations of the reverse diffusion process. Since the physics solvers require binary foam images and may fail to converge on highly noisy inputs, we rescale $\hat{\mathbf{x}}_0$ to [0, 1] and subsequently apply a projection to obtain a near-binary image. The projection function $\mathcal{P}$ is defined as:
\begin{equation}
\mathcal{P}(\hat{\mathbf{x}}_0) = 
\frac{
    \tanh\left( \frac{\eta}{2} \right)
    + \tanh\left[ \eta \left( \hat{\mathbf{x}}_0 - \frac{1}{2} \right) \right]
}{
    2 \tanh\left( \frac{\eta}{2} \right)
}
\end{equation}
where $\eta$ is a hyperparameter that controls the sharpness of the projection. A small $\eta$ results in a smooth transition between phases, while a large $\eta$ yields a more binary distribution. At the beginning of sampling, $\eta$ is set to a relatively small value to allow the model to freely explore the design space. As sampling proceeds, we gradually increase $\eta$ so that the final generated foam structure becomes fully binary. The material inherent property $\xi$ for each phase is then assigned according to a linear interpolation scheme:
\begin{equation}
    \xi(\hat{\mathbf{x}}_0) = \xi_0 + (\xi_1 - \xi_0)\mathcal{P} (\hat{\mathbf{x}}_0)
    \label{eq:property_interp}
\end{equation}
where $\xi_0$ and $\xi_1$ represent the inherent properties of two phases, and $\xi$ is the material property fed into the physics-based simulators during the sampling process.

The hyperparameter $s$ in Eq.~(\ref{eq:coreeq}) which controls the guidance intensity also needs to be carefully designed. Similar to the $\eta$ schedule, we use a low guidance intensity at the beginning of sampling to encourage exploration, and gradually increase it toward the end to strictly steer the generated structure toward the target properties.

The parameters used in our physics-guided diffusion model are listed in Table~\ref{tab:sde_params} for reference.

\begin{table}[H]
    \centering
    \caption{Parameters used in the physics-guided diffusion model.}
    \label{tab:sde_params}
    \begin{tabular}{ll}
        \toprule
        Parameters          & value            \\
        \midrule
        \text{Image resolution}                 & 64 \\
        \text{Channel}                      & 2 \\
        \text{Batch size}                    & 16 \\
        \text{Learning rate}             & \(2 \times 10^{-4}\) \\
        \text{Epochs}                  & 1000 \\
        \text{Sampling steps}         & 50 \\
        $t$                         & [0, 1] \\
        $\beta(t)$                   & 20$t$ \\
        $\eta (t)$                    & 5 $\times$ $2^{1-t}$ \\
        $s (t)$                   & 50$(1 - t)$ \\
        \bottomrule
    \end{tabular}
\end{table}

\section{Numerical examples}
\label{sec:case}

In this section, we present three  numerical examples to demonstrate the effectiveness and flexibility of the proposed physics-guided diffusion models for inverse design of disordered foams. All examples use the same well-trained diffusion model to achieve the design of thermal conductivity, load–displacement response, and energy absorption, respectively.

\subsection{Design of effective thermal conductivity}
\subsubsection{Problem formulation}

\begin{figure}[H]
    \centering
    \includegraphics[scale=0.5]{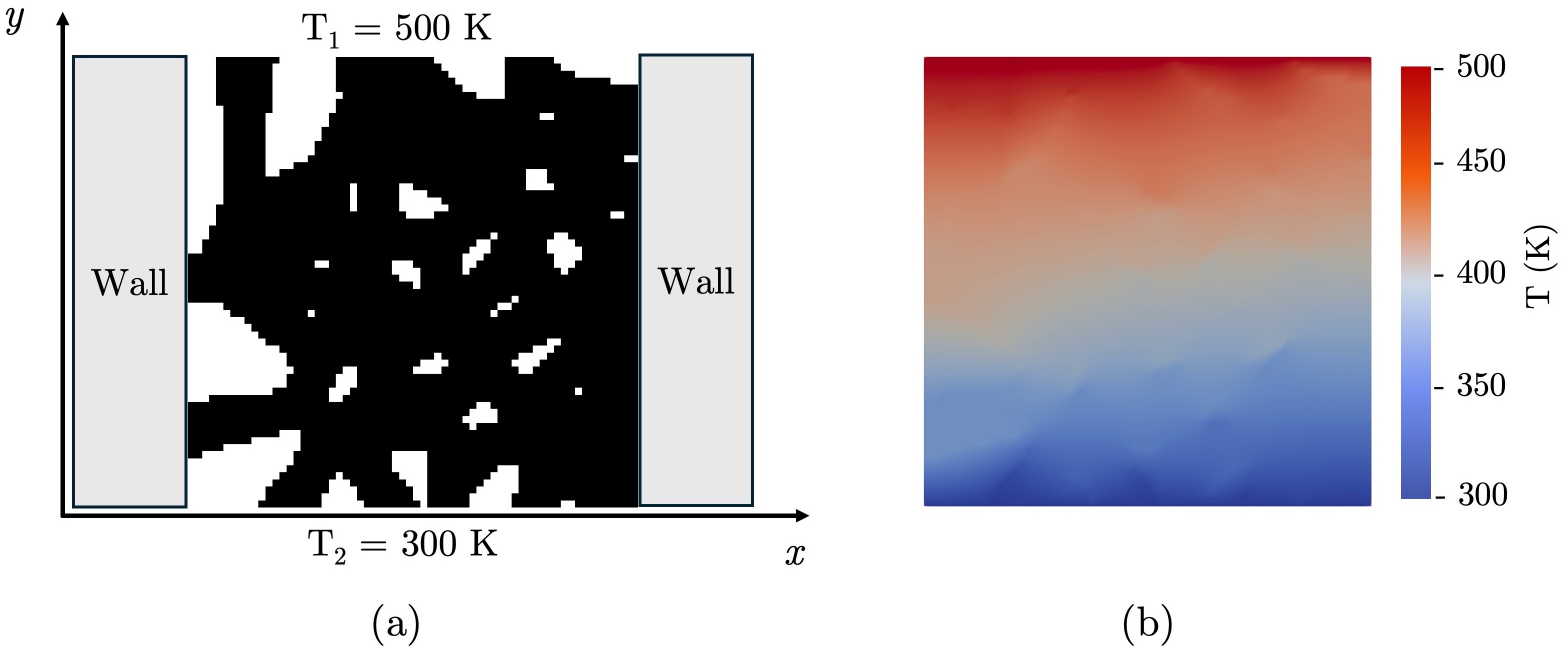}
    \caption{Formulation of the foam heat transfer problem: (a) Problem set-up; (b) Steady state temperature field.}
    \label{fig:k_formu}
\end{figure}

We consider a simple steady-state heat conduction problem in this subsection, and the aim is to design foam structures with a target thermal conductivity. The computational domain $\Omega$ represents the region with dimension 1 $\times$ 1~$\mathrm{m}^2$ between two walls, containing a disordered foam structure and voids filled with gas, as illustrated in Fig.~\ref{fig:k_formu}~(a). The top boundary is maintained at temperature $T_1 = 500~\mathrm{K}$, while the bottom boundary is held at $T_2 = 300~\mathrm{K}$. The left and right walls are assumed to be adiabatic.

The governing equation for steady-state heat conduction is given by:
\begin{equation}
-\nabla \cdot (k \nabla T) = 0
\end{equation}
with $k$ denotes the local thermal conductivity and $T$ is the temperature field. As a simplification, we treat both materials as having constant thermal conductivities, where 100~$\mathrm{W\cdot m^{-1} \cdot K^{-1}}$ for the solid phase and 1~$\mathrm{W\cdot m^{-1} \cdot K^{-1}}$ for the gas phase. The boundary conditions are specified as follows:
\begin{equation}
\begin{aligned}
T & = T_1 \quad &&\text{on } \Gamma_D^{\text{top}} \\
T & = T_2 \quad &&\text{on } \Gamma_D^{\text{bottom}} \\
-k\frac{\partial T}{\partial \mathbf{n}} &= 0 \quad &&\text{on } \Gamma_N^{\text{wall}}
\end{aligned}
\end{equation}
where $\mathbf{n}$ represents the outward unit normal vector to the boundary, $\Gamma_D^{\text{top}}$ and $\Gamma_D^{\text{bottom}}$ denote the top and bottom Dirichlet boundaries, $\Gamma_N^{\text{wall}}$ represents the adiabatic walls.


\texttt{JAX-FEM}~\cite{xue2023jax}, a differentiable GPU-accelerated solver based on FEM, is used here to solve for the steady-state temperature field, and an example steady-state solution is shown in Fig.~\ref{fig:k_formu}~(b). Then, the effective thermal conductivity of the structure is given as~\cite{mendes2013simple}:
\begin{equation}
k_{\text{eff}} = \frac{1}{\Delta T} \int_{L} k \left| \frac{\partial T}{\partial y} \right|_{y=\text{const}} dx
\end{equation}
where $k_{\text{eff}}$ is the effective thermal conductivity, $\Delta T$ is the temperature difference, $L$ is the side length perpendicular to heat flow, and $y$ and $x$ represent the directions parallel and perpendicular to the heat flow, respectively. In this work, the integration is performed over the top surface of foams.

To design disordered foams with specific effective thermal conductivity, we modify the sampling SDE of our well-trained diffusion model as:
\begin{equation}
    \mathrm{d}\mathbf{x} = \left[
        -\frac{1}{2}\beta(t) \mathbf{x} 
        - \beta(t) \left(
            -\frac{\boldsymbol{\epsilon}_\theta(\mathbf{x}_t, t)}{\sigma_t} 
            - s \underbrace{\nabla_{\mathbf{x}_t} \left\|k_\text{eff}^* - k_{\text{eff}}(\hat{\mathbf{x}}_0(\mathbf{x}_t)) \right\|^2}_\text{Physics guidance}
        \right)
    \right] \mathrm{d}t 
    + \sqrt{\beta(t)}\, \mathrm{d}\bar{\mathbf{w}}
\end{equation}
where $k_\text{eff}^*$ is the target effective thermal conductivity, $k_\text{eff}(\cdot)$ is the physics-based solver to perform the calculation of effective thermal conductivity on generated structures. Gradients are computed via customized reverse-mode AD, where specialized vector-Jacobian product (VJP) rules are derived from the implicit function theorem~\cite{rudin2021principles}.

\subsubsection{Results}

\begin{figure}[H]
\centering
\begin{subfigure}[b]{0.6\textwidth}
    \centering
    \includegraphics[width=\textwidth]{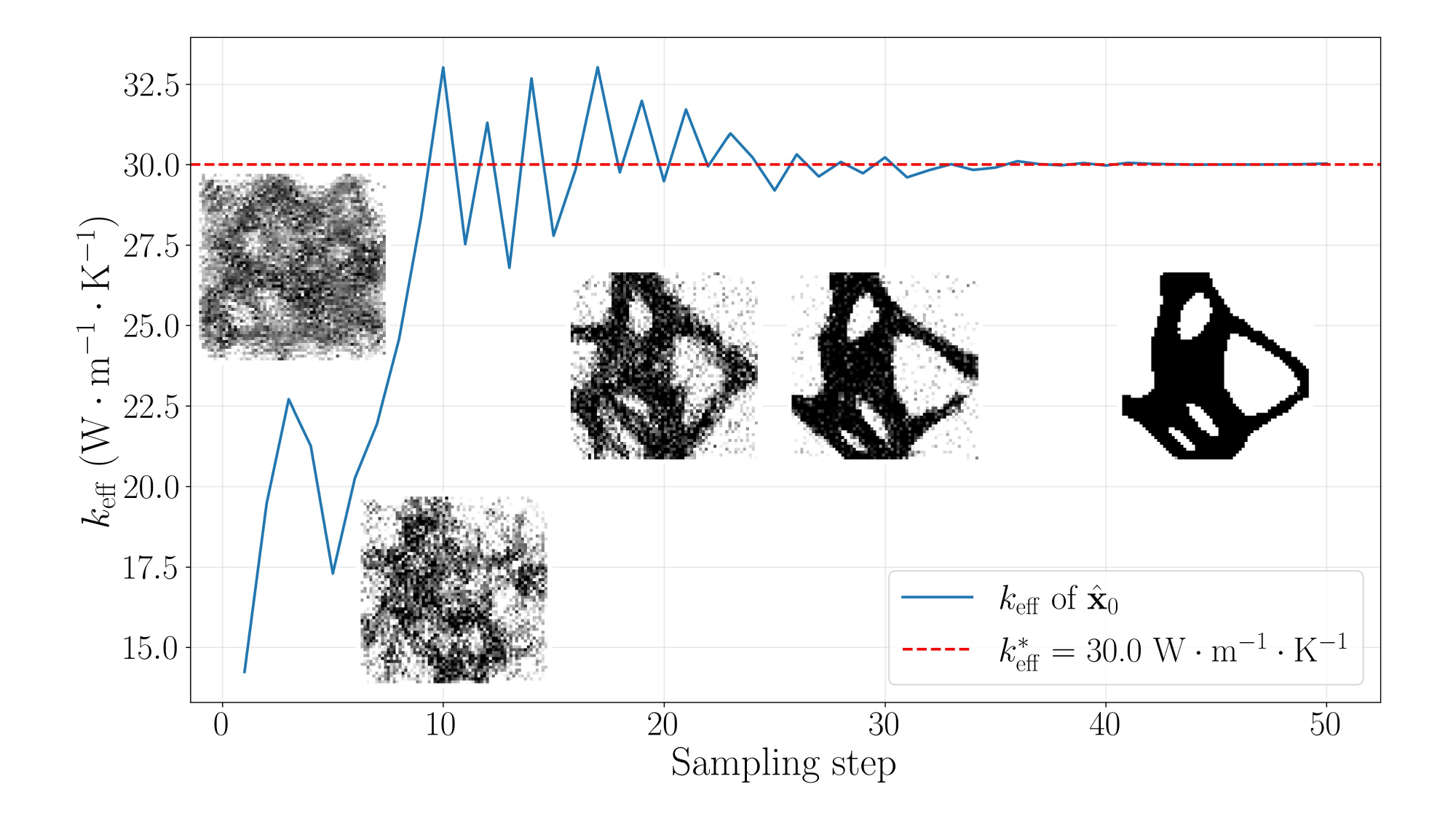}
    \caption{}
\end{subfigure}
\\[\baselineskip]
\begin{subfigure}[b]{0.55\textwidth}
    \centering
    \includegraphics[width=\textwidth]{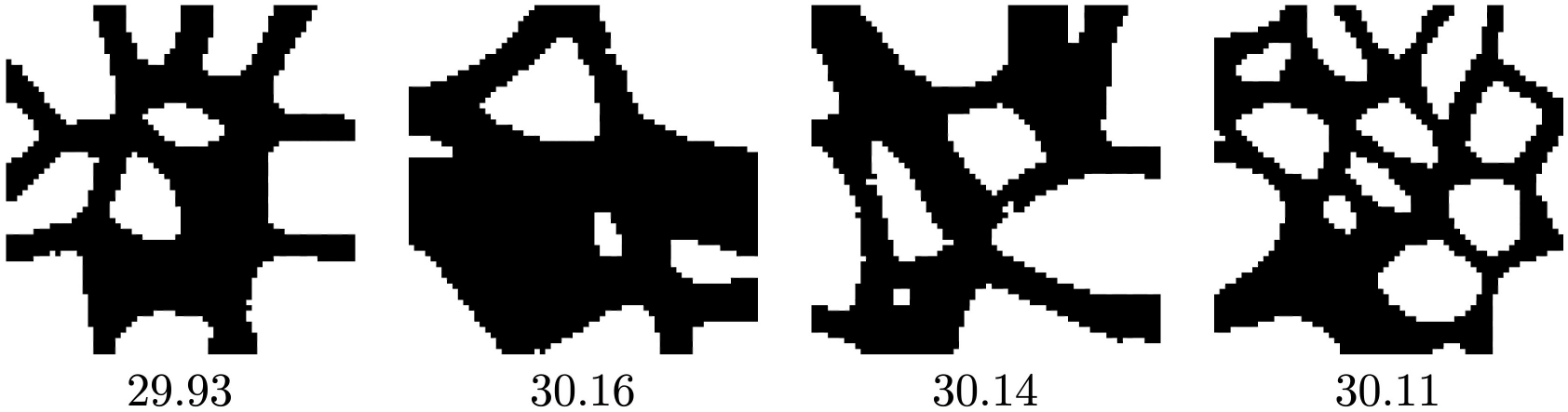}
    \caption{}
\end{subfigure}
\caption{Results of generated foams with target effective thermal conductivity $k^*_\text{eff} = 30$~$\mathrm{W{\cdot}m^{-1}{\cdot}K^{-1}}$: (a) Evolution of $k_\text{eff}$ and sampled structures; (b) Additional four samples with final $k_\text{eff}$ values captioned, and the unit is $\mathrm{W{\cdot}m^{-1}{\cdot}K^{-1}}$.}
\label{fig:res_therm}
\end{figure}

The generated foams with a targeted effective thermal conductivity \(k^*_\text{eff} = 30~\mathrm{W{\cdot}m^{-1}{\cdot}K^{-1}}\) are shown in Fig.~\ref{fig:res_therm}. Fig.~\ref{fig:res_therm}~(a) presents the convergence history of \(k_\text{eff}\) toward a target value of 30~\(\mathrm{W{\cdot}m^{-1}{\cdot}K^{-1}}\), with the blue solid line denoting the computed \(k_\text{eff}\) of the estimate \(\hat{\mathbf{x}}_0\) obtained from the physics-based solver at each sampling step and the red dashed line representing the target value. Several intermediate microstructure configurations are embedded along the curve, visually demonstrating the structural evolution from random noise to a clean and distinct disordered foam structure as generation progresses. Notably, \(k_\text{eff}\) exhibits substantial oscillations in the early steps, corresponding to rapid structural adaptations, before gradually approaching the target value with reduced variance as the structure stabilizes. The final $k_\text{eff}$ is 30.03~\(\mathrm{W{\cdot}m^{-1}{\cdot}K^{-1}}\) for this sample. 

Fig.~\ref{fig:res_therm}~(b) shows additional generated foam structures for the same target $k^*_{\mathrm{eff}} = 30~\mathrm{W\cdot m^{-1}\cdot K^{-1}}$. Under the framework of a measurement model $k_{\text{eff}}^* = k_{\text{eff}}(\mathbf{x}_0) + n$, these five independently generated samples exhibit small mean and variance of residuals, which are -0.074~$\mathrm{W{\cdot}m^{-1}{\cdot}K^{-1}}$ and $7.1 \times 10^{-3}~\mathrm{(W{\cdot}m^{-1}{\cdot}K^{-1})^2}$, respectively, corresponding to a high likelihood $p(k_{\text{eff}}^* | \mathbf{x}_0)$ and indicating reliable property control across the physics-guided generations.

Overall, these results demonstrate the efficacy and accuracy of the physics-guided sampling strategy in generating disordered foams with prescribed effective thermal conductivity.

\subsection{Control of load–displacement response}
\label{sec:MPM}
\subsubsection{Problem formulation}

We consider the compression of viscoelastic foams in this subsection, with the aim of designing foam structures that exhibit target load-displacement curves. The material considered in this task is flexible photosensitive resin. The conservation equations of mass and momentum are expressed as~\cite{hu2018moving}:
\begin{equation}
\begin{aligned}
\frac{D\rho}{Dt} + \rho \nabla \cdot \mathbf{v} &= 0 \\
\rho \frac{D\mathbf{v}}{Dt} &= \nabla \cdot \boldsymbol{\sigma} + \rho \mathbf{g}
\end{aligned}
\end{equation}
where \(\rho\) is the material density, \(\mathbf{v}\) is the velocity field, \(\mathbf{g}\) represents gravitational acceleration, and \(\frac{D(\cdot)}{Dt}\) denotes the material derivative.

\begin{figure}[H]
    \centering
    \includegraphics[width=0.7\linewidth]{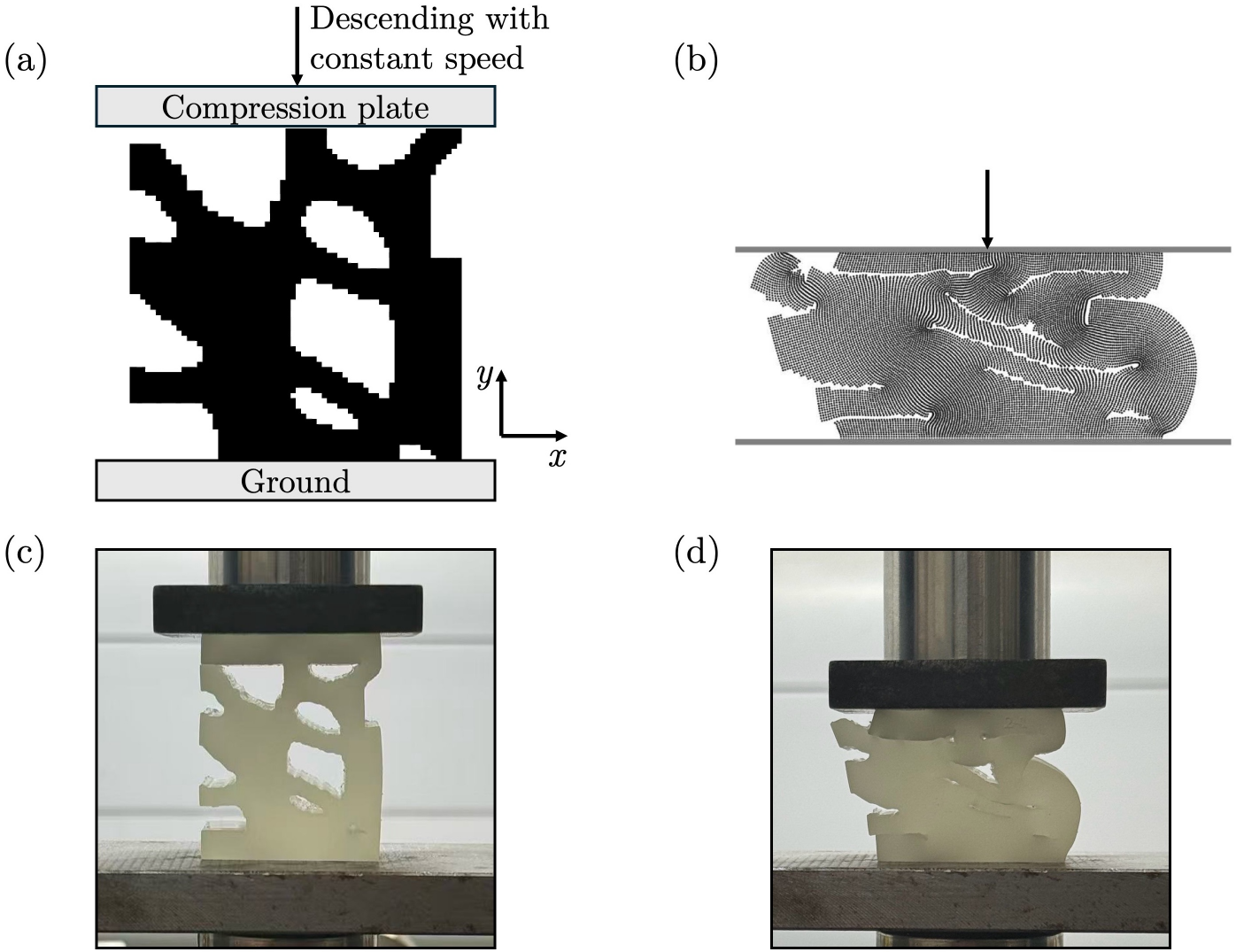}
    \caption{Formulation of the foam compression problem: (a) Problem set-up; (b) Compressed state in simulation; (c) Initial state in experiment; (d) Compressed state in experiment.}
    \label{fig:c_formu}
\end{figure}

As shown in Fig.~\ref{fig:c_formu}~(a), a foam sample is fixed on the ground, while a top plate moves downward at a constant velocity \(v_{\text{comp}}\). The boundary conditions are:
\begin{equation}
\begin{aligned}
\mathbf{v} &= \mathbf{0} && \text{on } \Gamma_D^{\text{ground}} \\
\mathbf{v} &= -v_{\text{comp}} \mathbf{e}_y && \text{on } \Gamma_D^{\text{plate}} 
\end{aligned}
\end{equation}
where $\mathbf{e}_y$ is the unit vector along the $y$ axis, \(\Gamma_D^{\text{ground}}\) and \(\Gamma_D^{\text{plate}}\) denote the ground and top plate boundaries, respectively.

The explicit moving least squares material point method (MLS-MPM)~\cite{hu2018moving} is applied to solve this problem. We implement MLS-MPM in \texttt{JAX} so that the solver is differentiable.  The material behavior is described by a finite-strain viscoelastic constitutive model composed of an equilibrium (hyperelastic) Neo-Hookean response and a non-equilibrium viscous overstress described by a generalized Maxwell model (Prony series)~\cite{simo1987fully, abaqus2011abaqus}. The damage-field gradient (DFG) partitioning frictional self-contact algorithm is additionally incorporated into our implementation~\cite{homel2017field, xiao2021dp}, enabling the automatic detection of potential contact surfaces and effectively preventing the merging of particles during the MPM simulation process. The final state of the foam in simulation is depicted in Fig.~\ref{fig:c_formu}~(b). More details of the MLS-MPM and constitutive model can be found in Appendix~\ref{apd:a}, while the frictional self-contact algorithm is described in Appendix~\ref{apd:b}.

The reaction force on the top plate is computed from the internal forces of top boundary nodes, which is given as $F = -\sum_{i \in N_{\text{top}}} \mathbf{f}_i \cdot \mathbf{e}_y$, where \(N_{\text{top}}\) is the set of top boundary nodes, and $\mathbf{f}_i$ is the internal force of grid node $i$. The load-displacement curve plots load against displacement \(u = v_{\text{comp}} \cdot t\). Experiments are performed to calibrate our simulation as shown in Fig.~\ref{fig:c_formu}~(c) and (d), corresponding to the initial and compressed states of the foam resin, respectively. A 3D printer for photosensitive resin is used to fabricate the foams. To facilitate the realization of no-slip boundary conditions on both the ground and the compression plate in the experiments, additional resin layers are applied to the top and bottom surfaces of the foam structure. These resin pads are also incorporated in our simulations to ensure consistency between the experimental setup and the numerical model.

To design foam structures with desired load-displacement curves, we use $n=100$ discretized points to describe the curve, and evaluate the match degree between the target curve and the current curve of $\hat{\mathbf{x}}_0$ with mean square error (MSE). The physics-guided sampling equation is given as:
\begin{equation}
    \mathrm{d}\mathbf{x} = \left[
        -\frac{1}{2}\beta(t) \mathbf{x} 
        - \beta(t) \left(
            -\frac{\boldsymbol{\epsilon}_\theta(\mathbf{x}_t, t)}{\sigma_t} 
            - s \underbrace{\nabla_{\mathbf{x}_t} \frac{1}{n}\sum_{j=1}^{n} \left\|c_j^* - c_j (\hat{\mathbf{x}}_0(\mathbf{x}_t)) \right\|^2}_\text{Physics guidance}
        \right)
    \right] \mathrm{d}t 
    + \sqrt{\beta(t)}\, \mathrm{d}\bar{\mathbf{w}}
\end{equation}
where $c^*$ is the target load-displacement curve, $c (\cdot)$ denotes the physics-based solver to simulate the compression response of given structures, and $j$ is the index of points. In this task, \(c_j\) is equal to the load value \(F_j\). Gradients are computed by backpropagation through the explicit MPM time-stepping scheme using the reverse-mode AD.

\subsubsection{Results}

\begin{figure}[H]
\centering
\begin{subfigure}[b]{0.66\textwidth}
    \centering
    \includegraphics[width=\textwidth]{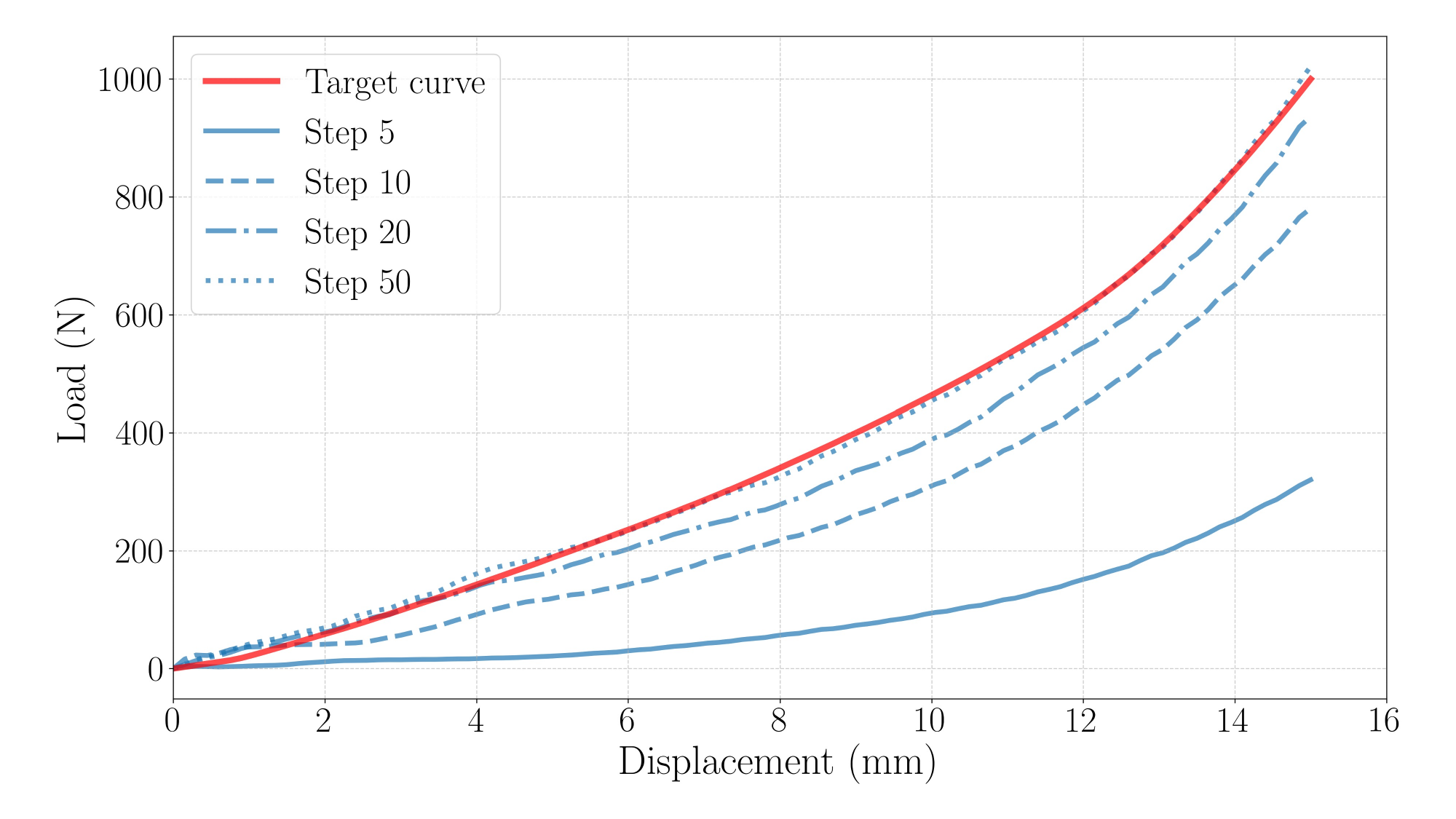}
    \caption{}
\end{subfigure}
\\[\baselineskip]
\begin{subfigure}[b]{0.7\textwidth}
    \centering
    \includegraphics[width=\textwidth]{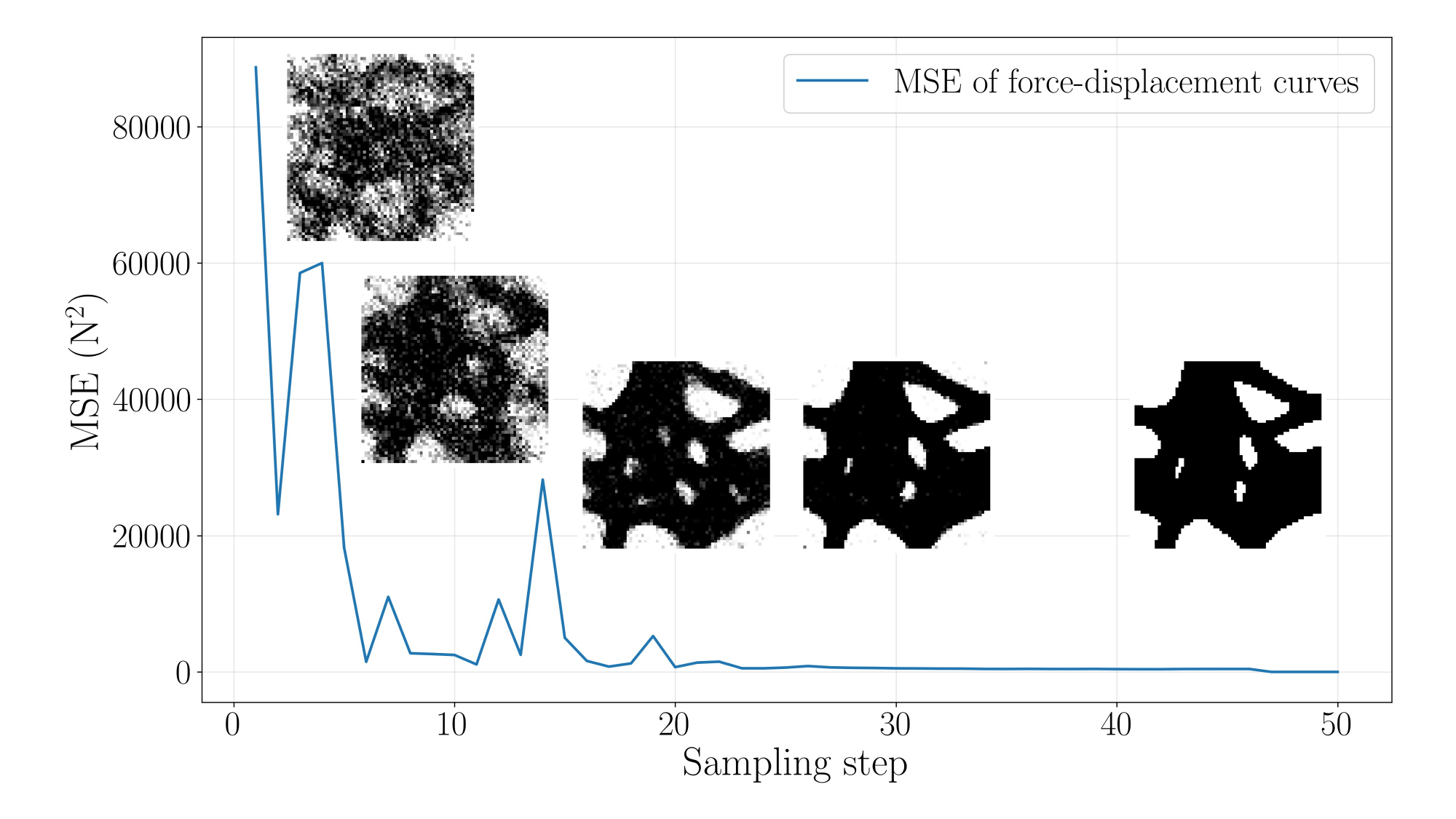}
    \caption{}
\end{subfigure}
\caption{Evolution of the (a) simulated load-displacement curves of representative sampling steps and (b) MSE and generated structures during the guided sampling process.}
\label{fig:c_evo}
\end{figure}

The first example demonstrates the generation of a foam structure with a prescribed hyperelastic target curve. Fig.~\ref{fig:c_evo}~(a) presents the corresponding target and simulated load-displacement curves at representative sampling steps. Starting from a significantly softer response at step 5, the predicted curves progressively approach the target curve. Fig.~\ref{fig:c_evo}~(b) shows the evolution of MSEs of the load-displacement responses during the sampling process. The MSE decreases dramatically from an initial value of approximately 80,000~N$^2$ in the early iterations, exhibiting several peaks corresponding to significant structural adjustments (as illustrated by the vague foam morphologies at early stages). After step 20, the loss stabilizes below 5,000~N$^2$, indicating convergence to a stable structure. The inset images reveal the gradual structural evolution from the initial random configurations into a foam structure with optimized void distributions. 

\begin{figure}[H]
\centering
\begin{subfigure}[b]{0.8\textwidth}
    \centering
    \includegraphics[width=\textwidth]{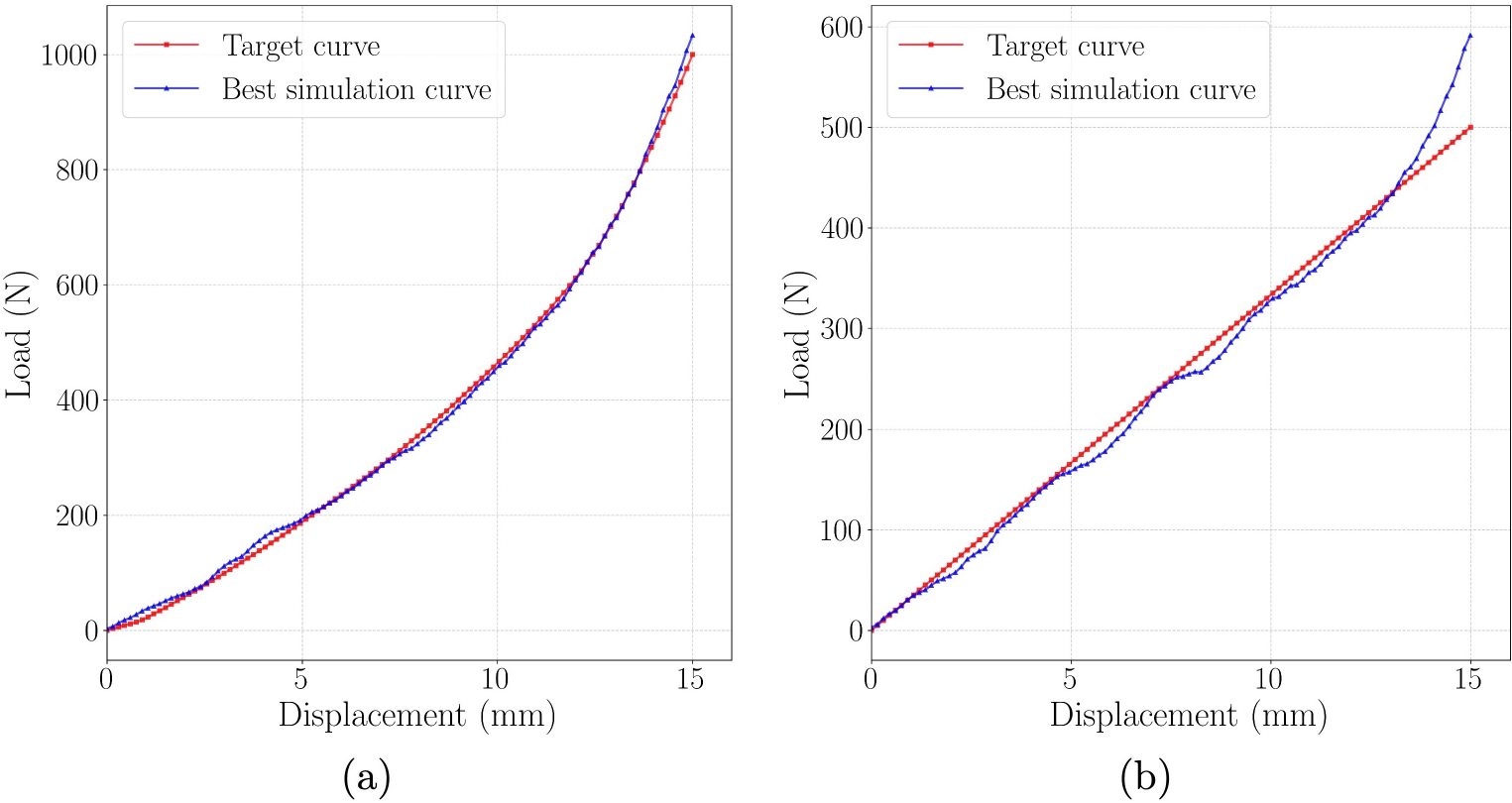}
\end{subfigure}
\\[\baselineskip]
\begin{subfigure}[b]{0.8\textwidth}
    \centering
    \includegraphics[width=\textwidth]{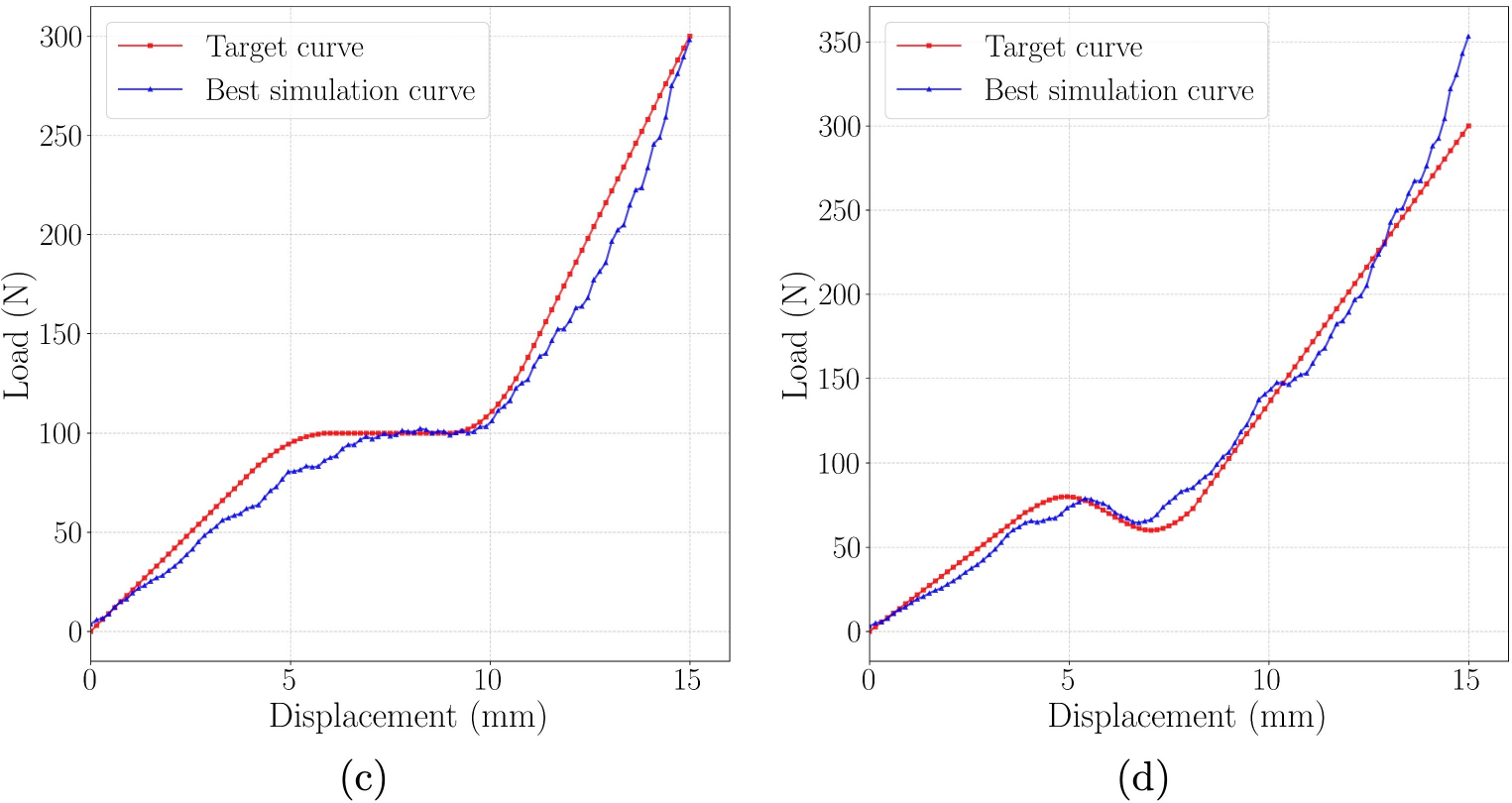}
\end{subfigure}
\caption{Best-matching simulated load–displacement curve for each design target, selected from five generated samples.}
\label{fig:c_4cases}
\end{figure}

To systematically evaluate the inverse design capability of the proposed framework across a broad range of mechanical behaviors, four representative target load-displacement curves are designed, as shown in Fig.~\ref{fig:c_4cases}: (a) hyperelastic behavior, reaching 1000~N at 15~mm displacement; (b) linear response with a maximum load of 500~N; (c) classical foam response~\cite{liu2014numerical} featuring a pronounced plateau at 100~N between 5 and 10~mm; and (d) complex non-monotonic behavior with extreme buckling and densification. For each case, we generate five different foams and plot the best simulation results on the corresponding figures. From Fig.~\ref{fig:c_4cases}, it can be observed that the best simulation curves (blue) closely match the targets (red) in cases (a) and (b), while minor discrepancies exist in cases (c) and (d), particularly during the buckling and densification processes. 

\begin{figure}[H]
\centering
\begin{subfigure}[b]{0.8\textwidth}
    \centering
    \includegraphics[width=\textwidth]{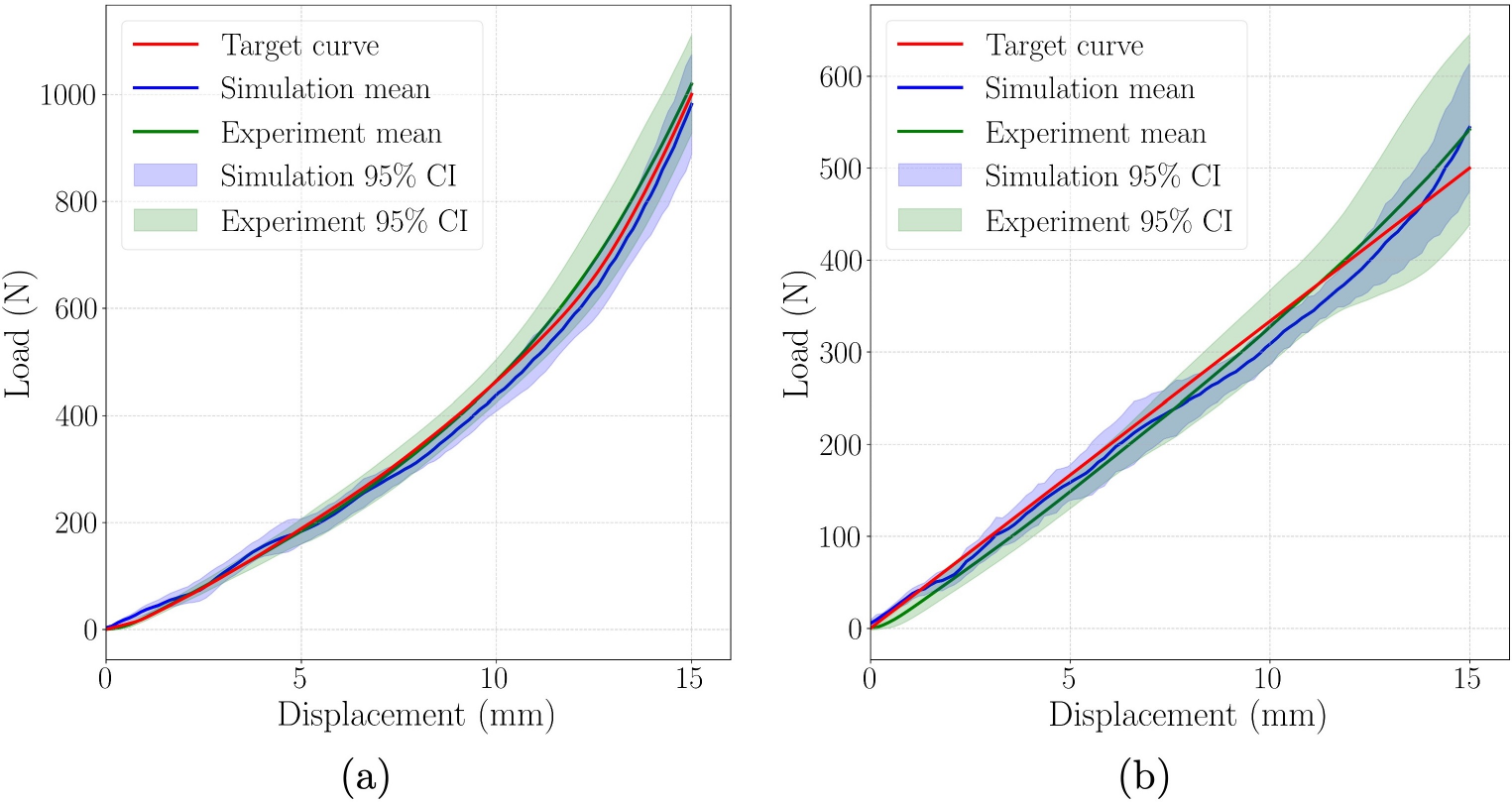}
\end{subfigure}
\\[\baselineskip]
\begin{subfigure}[b]{0.8\textwidth}
    \centering
    \includegraphics[width=\textwidth]{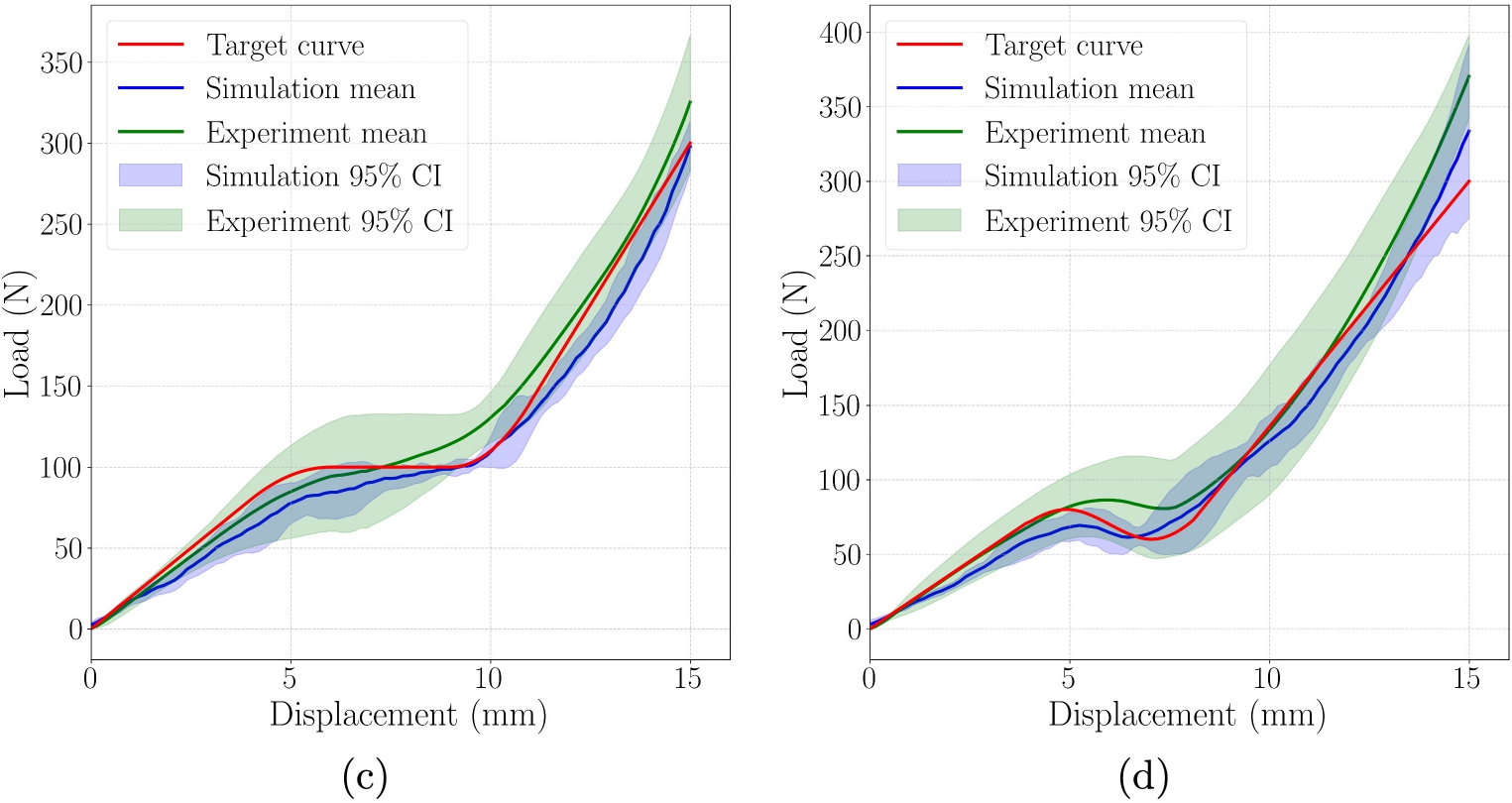}
\end{subfigure}
\caption{Means and 95$\%$ confidence intervals of the load-displacement curves of the five generated samples for each design target.}
\label{fig:c_div}
\end{figure}

For all samples generated in this task, we record the simulated responses and also conduct corresponding compression experiments to obtain the experimental curves. The means and 95\% confidence intervals (CI) of the simulated and experimental load-displacement curves of the five generated foams in each case are shown in Fig.~\ref{fig:c_div}. Notably, all five generated samples in each case demonstrate good agreement with the target curves (red lines), as evidenced by the narrow CIs (blue and green bands) that closely envelope the targets across the entire displacement range. For the simple target curves in cases (a) and (b), the generated foams exhibit excellent matching with minimal deviations, with both simulation and experimental results tightly clustered around the targets. For the more complex target curves in cases (c) and (d), although the matching accuracy is slightly reduced compared to cases (a) and (b), the overall trends of all generated samples remain consistent with the targets, successfully capturing the characteristic features such as plateau stress levels and densification onset. 

These results demonstrate that the proposed physics-guided sampling method can reliably generate diverse structures that consistently satisfy the prescribed mechanical specifications, including challenging non-monotonic responses.

\subsection{Maximization of energy absorption involving fractures}
\label{sec:pf}
\subsubsection{Problem formulation}

The energy absorption capability of foams enduring fracture is studied in this subsection, and the aim is to design foam structures with maximal absorbed energy given the specified volume fraction. Within the phase field brittle fracture framework~\cite{miehe2010phase, xue2021mapped}, the fracture evolution can be characterized through minimization of the total energy functional with respect to the displacement field $\mathbf{u}$ and the phase field variable $\zeta$:
\begin{align}
    \Pi(\mathbf{u}, \zeta) = \int_{\Omega} \psi(\mathbf{\boldsymbol{\varepsilon}}(\mathbf{u}), \zeta)\, \mathrm{d}\Omega + \int_{\Omega}g_c\gamma(\zeta, \nabla \zeta)\, \mathrm{d}\Omega  - \int_{\Omega}\mathbf{b}\cdot\mathbf{u}\, \mathrm{d}\Omega - \int_{\Gamma_N}\mathbf{t}\cdot\mathbf{u}\, \mathrm{d}\Gamma
\end{align}
where $\psi$ is the bulk elastic energy, $g_c$ denotes the critical energy release rate in the sense of Griffith, and $\gamma$ represents the regularized crack surface density function per unit volume.

\begin{figure}[H]
    \centering
    \includegraphics[width=0.7\linewidth]{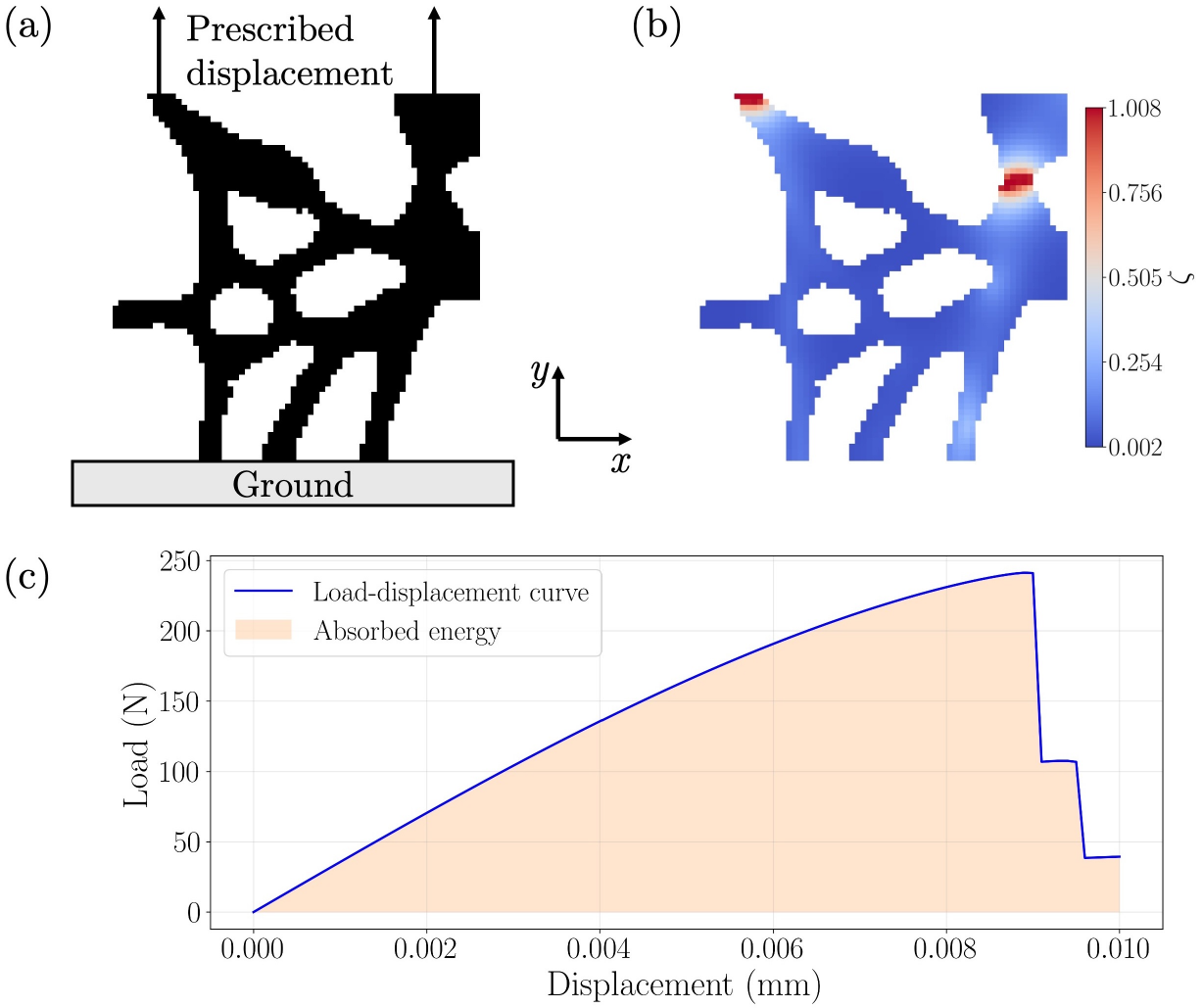}
    \caption{Formulation of the foam brittle fracture problem: (a) Problem set-up; (b) Solution of the phase field variable; (c) Load-displacement curve and absorbed energy.}
    \label{fig:f_formu}
\end{figure}

As shown in Fig.~\ref{fig:f_formu}~(a), a foam is fixed on the ground and subjected to a prescribed displacement, and the fracture behavior of the foam is analyzed. The governing equations and associated boundary conditions derived from minimizing the total energy functional are expressed as:
\begin{equation}
\begin{aligned}
    \nabla \cdot \boldsymbol{\sigma} + \mathbf{b} &= \mathbf{0} && \text{in}\;\Omega \\
    \frac{g_c}{l}\left( \zeta - l^2 \Delta \zeta \right) &= 2(1-\zeta) \mathcal{H} && \text{in}\;\Omega \\
    \mathbf{u} &= \mathbf{u}_D && \text{on}\;\Gamma_D \\
    \boldsymbol{\sigma} \cdot \mathbf{n} &= \mathbf{t} && \text{on}\;\Gamma_N \\
    \nabla \zeta \cdot \mathbf{n} &= 0 && \text{on}\;\Gamma 
\end{aligned}
\end{equation}
where $\boldsymbol{\sigma} = \frac{\partial \psi}{\partial \mathbf{\boldsymbol{\varepsilon}}}=g(\zeta)\boldsymbol{\sigma}^+ + \boldsymbol{\sigma}^-$, with the decomposition scheme from Miehe et al.~\cite{miehe2010phase}:

\begin{align}
\boldsymbol{\sigma}^{\pm} = \frac{\partial \psi^{\pm}}{\partial \mathbf{\boldsymbol{\varepsilon}}} = \lambda \langle\mathrm{tr}(\mathbf{\boldsymbol{\varepsilon}})\rangle_{\pm} \mathbf{I} + 2\mu \mathbf{\boldsymbol{\varepsilon}}_{\pm}
\end{align}

The history field is introduced to ensure irreversibility of crack growth during unloading cycles:
\begin{align}
 \mathcal{H}(\mathbf{x}, t)= \max_{q\in[0, t]}\psi_+(\mathbf{\boldsymbol{\varepsilon}}(\mathbf{x}, q))
\end{align}

More information about the phase field brittle fracture model and the corresponding model parameters used in this work can be found in Appendix~\ref{apd:c}. 

\texttt{JAX-FEM}~\cite{xue2023jax} is used here to solve this problem, and an example solution of the phase field variable and load-displacement curve is shown in Fig.~\ref{fig:f_formu}~(b) and (c). The absorbed energy is defined as the area under the load–displacement curve up to the displacement of 0.01~mm. 

The phase field fracture problem can be solved with a monolithic scheme that simultaneously solves for the displacement and phase field variables, or a staggered scheme that alternately solves for the two fields. The monolithic scheme can be more accurate and efficient, but it may suffer from convergence issues. Therefore, the staggered scheme is usually adopted in practice for its robustness and stability, which can also provide acceptable accuracy with strict convergence criteria. 

In this work, we develop a hybrid strategy (staggered scheme for the forward problem; monolithic scheme for the inverse problem) for the phase field fracture analysis. 
This hybrid strategy leverages the robustness of the staggered scheme for forward solutions while avoiding the direct differentiation through the staggered iterations, which can be computationally expensive and memory-intensive. By treating the staggered solution as the equilibrium state of the monolithic residuals, we can efficiently compute the gradients using reverse-mode AD, thus achieving a stable and efficient differentiation process for the phase field fracture analysis. 

The detailed algorithm is summarized in Algorithm~\ref{alg:pf_hybrid}. 
With the input parameters $\mathbf{x}$, the forward problem is solved using the staggered scheme, where the displacement field is solved first with a fixed phase field, followed by the update of the phase field with the updated displacement. This process is repeated until reaching the relative tolerance of $\epsilon=10^{-5}$ for both fields. After that, the objective function $f$ is evaluated. The obtained forward solution $\mathbf{y}_n = [\mathbf{u}_n;\,\zeta_n]$ at each loading step $n$ is then treated as the equilibrium state of the monolithic residuals $\mathbf{R}_n$. 
In the backward pass, we leverage the reverse-mode AD to compute the gradients of $f$ with respect to $\mathbf{x}$. A customized VJP rule is derived based on the monolithic residuals, which involves solving an adjoint equation to obtain the adjoint variable $\boldsymbol{\mu}_n$, and then accumulating the gradients of $\mathbf{x}$ and continuing to propagate until the initial step.

The probability  $p(E | \hat{\mathbf{x}}_0)$ is modeled by an energy-based model (EBM) through Boltzmann distribution~\cite{xie2016theory, gao2020learning}:

\begin{equation}
    p(E | \hat{\mathbf{x}}_0) \propto \exp\!\big(E(\mathbf{\hat{\mathbf{x}}_0})\big)
\end{equation}

For the task of maximizing the absorbed energy of generated foams subject to a volume-fraction constraint, the physics-guided sampling equation is expressed as:
\begin{equation}
    \mathrm{d}\mathbf{x} = \left\{
        -\frac{1}{2}\beta(t) \mathbf{x} 
        - \beta(t) \left[
            -\frac{\boldsymbol{\epsilon}_\theta(\mathbf{x}_t, t)}{\sigma_t} 
            + s \underbrace{\left(- m \nabla_{\mathbf{x}_t} \left\|\phi^* - \phi(\hat{\mathbf{x}}_0(\mathbf{x}_t)) \right\|^2
            + \nabla_{\mathbf{x}_t} E(\hat{\mathbf{x}}_0(\mathbf{x}_t) \right)}_\text{Physics guidance}
        \right]
    \right\} \mathrm{d}t 
    + \sqrt{\beta(t)}\, \mathrm{d}\bar{\mathbf{w}}
\label{eq:pf_sample}
\end{equation}
where $\phi^*$ is the limitation of volume fraction, $\phi (\cdot)$ is the solver for detecting the volume fraction of the input structure, $E(\cdot)$ is the solver for absorbed energy, and $m$ is a scaling factor that adjusts the magnitude of the volume-guidance term to be comparable to that of the energy-guidance term. Empirically, we find that the volume-guidance term should be set slightly stronger than the energy-guidance term to ensure that the generated structures satisfy the volume-fraction constraint. Accordingly, we set the scaling factor to \(m = 5\times 10^{-5}\) in this task.

\begin{algorithm}[H]
\caption{Hybrid strategy for phase field fracture analysis}
\label{alg:pf_hybrid}
\begin{algorithmic}[1]
\State \textbf{Input:} Parameters $\mathbf{x}$
\State \textbf{Forward pass (Staggered scheme):}
\State Initialize $\mathbf{u}_0$, $\zeta_0$, and $\mathcal{H}_0$
\For{$n = 1$ to $N$}
    \State Initialize $k=0$, $\mathbf{u}_n^{(0)}=\mathbf{u}_{n-1}$, $\zeta_n^{(0)}=\zeta_{n-1}$, $\mathcal{H}_n^{(0)}=\mathcal{H}_{n-1}$
    \Repeat
    \State Fix $\zeta_n^{(k)}$, solve: $\mathbf{R}_u(\mathbf{u}_n^{(k+1)},\zeta_n^{(k)};\mathbf{x})=\mathbf{0}$
    \State Compute $\mathcal{H}_n^{(k+1)}$
    \State Fix $\mathbf{u}_n^{(k+1)}$, solve: $\mathbf{R}_\zeta(\mathbf{u}_n^{(k+1)},\zeta_n^{(k+1)},\mathcal{H}_n^{(k+1)};\mathbf{x})=\mathbf{0}$
    \State $k \gets k+1$
\Until{$\dfrac{\|\mathbf{u}_n^{(k)}-\mathbf{u}_n^{(k-1)}\|}{\|\mathbf{u}_n^{(k)}\|}<\epsilon$ \textbf{ and } $\dfrac{\|\zeta_n^{(k)}-\zeta_n^{(k-1)}\|}{\|\zeta_n^{(k)}\|}<\epsilon$}
\State Set $\mathbf{u}_n=\mathbf{u}_n^{(k)}$, $\zeta_n=\zeta_n^{(k)}$, $\mathcal{H}_n=\mathcal{H}_n^{(k)}$
\EndFor
\State Compute objective $f(\{\mathbf{u}_n,\zeta_n\}_{n=0}^N;\mathbf{x})$
\State \textbf{Backward pass (Monolithic scheme):}
\State Initialize seed $\bar{f}=1$, accumulator $\bar{\mathbf{x}}=\mathbf{0}$
\State Compute $\bar{\mathbf{y}}_n = (\dfrac{\partial f}{\partial \mathbf{y}_n})^\top\bar{f}$ for all $n$
\For{$n = N$ down to $1$}
    \State Assemble $\mathbf{y}_n=[\mathbf{u}_n;\,\zeta_n]$, $\mathbf{R}_n=[\mathbf{R}_u(\mathbf{y}_n;\mathbf{x});\,\mathbf{R}_\zeta(\mathbf{y}_n,\mathcal{H}_n;\mathbf{x})]$
    \State Solve adjoint equation: $\left(\dfrac{\partial \mathbf{R}_n}{\partial \mathbf{y}_n}\right)^\top \boldsymbol{\mu}_n = -\,\bar{\mathbf{y}}_n$
    \State Accumulate: $\bar{\mathbf{x}} \mathrel{+}= (\dfrac{\partial \mathbf{R}_n}{\partial \mathbf{x}})^\top\boldsymbol{\mu}_n$
    \State Propagate: $\bar{\mathbf{y}}_{n-1} \mathrel{+}= (\dfrac{\partial \mathbf{R}_n}{\partial \mathbf{y}_{n-1}})^\top\boldsymbol{\mu}_n$
\EndFor
\State \textbf{Output:} $\nabla_{\mathbf{x}} f = \bar{\mathbf{x}}$
\end{algorithmic}
\end{algorithm}

\subsubsection{Results}

\begin{figure}[H]
\centering
\begin{subfigure}[b]{0.7\textwidth}
    \centering
    \includegraphics[width=\textwidth]{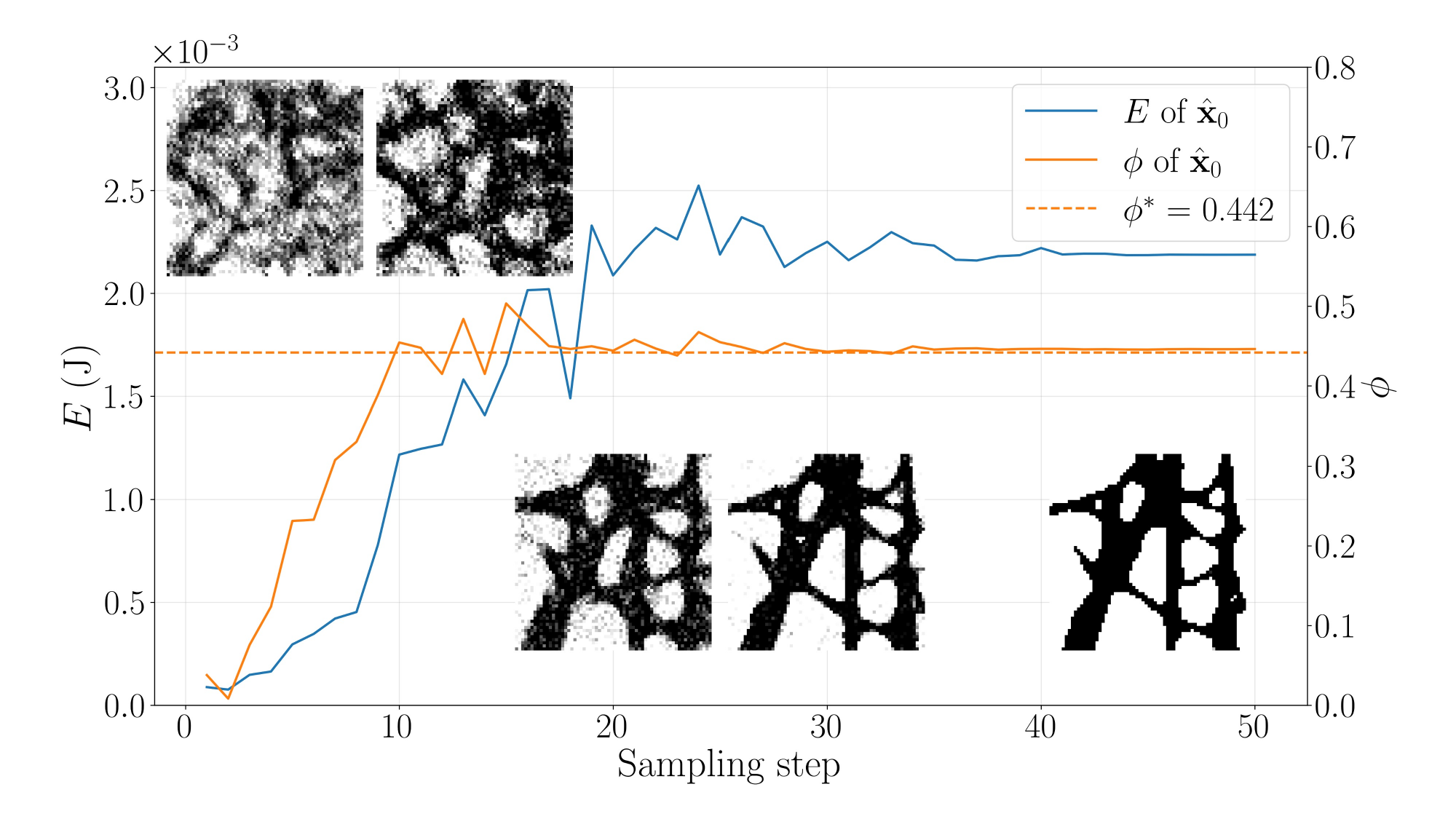}
    \caption{}
\end{subfigure}
\\[\baselineskip]
\begin{subfigure}[b]{0.65\textwidth}
    \centering
    \includegraphics[width=\textwidth]{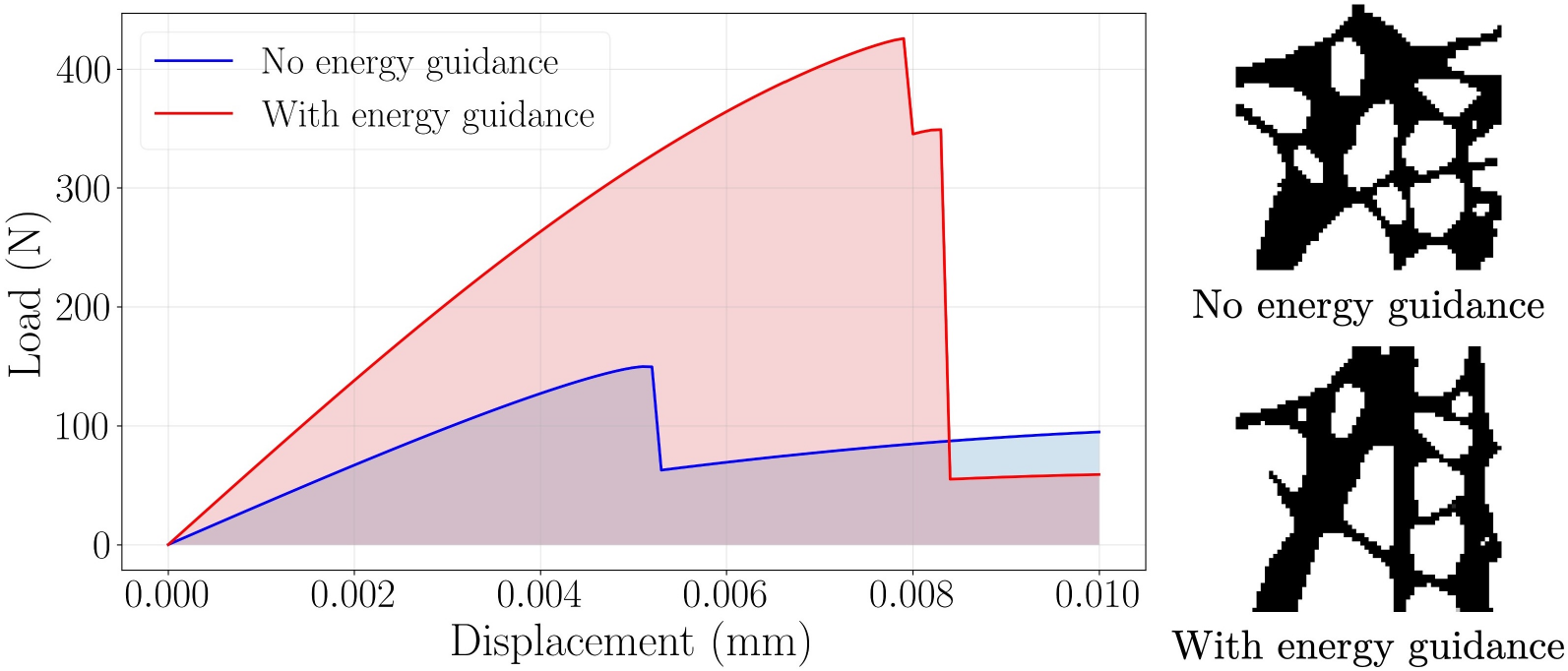}
    \caption{}
\end{subfigure}
\\[\baselineskip]
\begin{subfigure}[b]{0.65\textwidth}
    \centering
    \includegraphics[width=\textwidth]{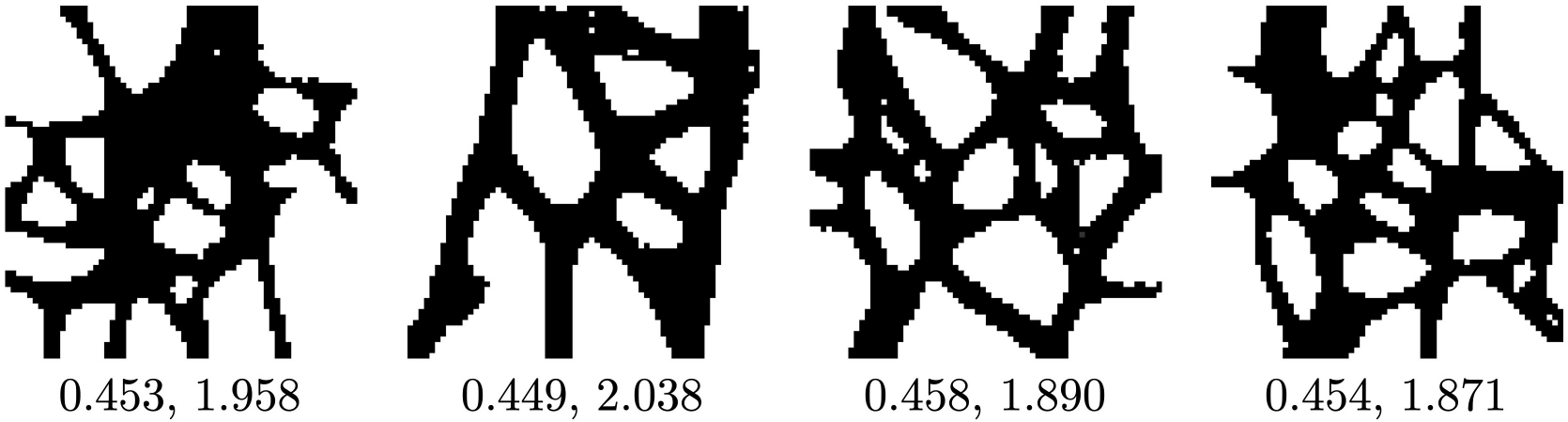}
    \caption{}
\end{subfigure}
\caption{Results of generated foams with maximized absorbed energy under a volume-fraction constraint $\phi^* = 0.442$: (a) Evolution of $E$, $\phi$ and sampled structures; (b) Comparison of load–displacement curves and structures of foams generated with and without energy guidance, using the same random seed; (c) Additional four samples with final $\phi$ and $E$ values captioned. The unit for energy is $10^{-3}$~J.}
\label{fig:e_res}
\end{figure}

We first randomly generate 10 foam structures without any guidance and evaluate their fracture energies via phase field simulations. Across these samples, the mean volume fraction is $0.442$ and the mean absorbed energy is $0.882 \times 10^{-3}~\mathrm{J}$. We then set the volume-fraction constraint $\phi^*$ to the mean value obtained from these ten structures and perform guided sampling according to Eq.~(\ref{eq:pf_sample}). The results are shown in Fig.~\ref{fig:e_res}. 

In Fig.~\ref{fig:e_res}~(a), we plot the evolution of the absorbed energy and the volume fraction along the sampling trajectory, together with the corresponding generated structures. With joint guidance in this case, both curves gradually level off in the final sampling stages, indicating that the absorbed-energy and volume-fraction guidance converge in a balanced manner. The final volume fraction of this sample is $0.446$, which is close to the prescribed volumetric constraint $0.442$. The final absorbed energy is $2.188 \times 10^{-3}~\mathrm{J}$, which is significantly higher than the average absorbed energy $0.882 \times 10^{-3}~\mathrm{J}$. In Fig.~\ref{fig:e_res}~(b), we also present the load-displacement curves and structures of foams generated with and without energy guidance, while the volume-fraction constraint is kept. For a fair comparison, the random seed is the same as that used in Fig.~\ref{fig:e_res}~(a). Without energy guidance, the absorbed energy of the generated structure is \(0.824 \times 10^{-3}~\mathrm{J}\), while with energy guidance, it increases by 166$\%$ to the value reported above.

Additional four foam structures under the same guidance of limiting the volume fraction and maximizing absorbed energy are shown in Fig.~\ref{fig:e_res}~(c). The mean volume fraction across all five samples is $0.452$, which closely matches the volume-fraction constraint. The average absorbed energy is $1.989 \times 10^{-3}~\mathrm{J}$, exceeding twice that of the samples generated without guidance. The guided samples tend to exhibit slightly larger volume fractions than the volume limitation, because adding material is encouraged by the energy-guidance term but penalized by the volume constraint. The resulting designs therefore settle at an intermediate compromise between these competing objectives.

These results demonstrate that the proposed physics-guided sampling strategy can generate foams with high energy absorption performance while satisfying a prescribed volume-fraction constraint. Moreover, the results validate the capability of our method to address multi-constraint inverse design problems.

\section{Discussion}
\label{sec:disc}

Our results in Sec.~\ref{sec:case} demonstrate that the proposed physics-guided diffusion models exhibit clear advantages in the inverse design of disordered metamaterials:  

\begin{enumerate}
    \item Physics consistency:
    By integrating differentiable physics solvers into the generative process, the approach enables direct steering of the sampling trajectory towards structures that strictly satisfy desired properties, such as specific thermal conductivity, load–displacement response, or energy absorption shown in the diverse case studies in this work. 

    \item Label-free and solver-decoupled training:
    The proposed diffusion framework can be trained on unlabeled datasets, avoiding the expensive generation of property labels typically required by conditional generative models. Moreover, by decoupling the physics solver from training and using it only for sampling, the approach avoids repeated solver evaluations and backpropagation across batches and diffusion steps, leading to substantially improved computational efficiency and training stability.    
    
    \item Flexibility: 
    The physics-guided approach can be flexibly adapted to a variety of inverse problems simply by modifying the guidance term, as evidenced by its effective performance across all three different design objectives in this work.
\end{enumerate}

However, some limitations still remain and need further investigations:

\begin{enumerate}
    \item Limited data distribution:
    Although the framework does not require labeled data, its generative prior is shaped by the training distribution and thus primarily explores patterns and the associated property ranges represented in the dataset. For example, a model trained on foams cannot produce layered or periodic designs. This motivates future work on broader and more diverse data curation to further expand the reachable design space.

    \item Parameter reliance:
    Performance depends on the choice of physics-guidance intensity. Overly aggressive guidance may slow convergence or introduce unreasonable designs. This suggests the need for automated parameter-calibration strategies that make the method more robust and user-friendly.

    \item Physics-solver reliance:
    The approach relies on the efficiency and fidelity of differentiable physics solvers, and solver evaluations can become the dominant cost for complex physics or large-scale 3D systems. This provides a direction for developing faster and higher-fidelity physics solvers or learned surrogates.
\end{enumerate}

\section{Conclusion}
\label{sec:conclusion}

In this work, we propose physics-guided diffusion models that leverage differentiable physics-based solvers to address inverse design problems of disordered metamaterials, focusing on closed-cell foam materials as the primary subject. We begin by generating a dataset of foam structures using Voronoi tessellation, which is then used to train a score-based diffusion model formulated through SDEs. Inspired by the concept of classifier guidance, we further modify the sampling process by incorporating physics-based guidance into the SDEs, where the guidance is evaluated via differentiable physics solvers. To validate the proposed approach, we provide three numerical case studies: (1) the design of foams with target effective thermal conductivity, (2) tailoring the load-displacement response, and (3) maximizing energy absorption under a given volume-fraction constraint. The results consistently demonstrate the efficacy of physics-guided diffusion models in solving challenging inverse design tasks.

In contrast to conventional generative models that require large labeled datasets or physics residuals in model training, the proposed physics-guided diffusion framework can be trained only with unlabeled data, significantly reducing the need for annotated samples and computational costs. Moreover, the framework offers flexible adaptation to various inverse problems simply by modifying the gradient term in the sampling equation. The diversity of the presented numerical examples highlights the versatility and practicality of the physics-guided sampling technique.

\section{Acknowledgments}
This work was financially supported by National Natural Science Foundation of China (YSF 1240020027, JRS N$\_$HKUST687/25), Research Grants Council of Hong Kong (ECS 26205024, CRF C7085-24G) and Innovation and Technology Fund (ITS/032/23).

\bibliographystyle{unsrt}
\bibliography{refs}

\newpage
\section*{Appendix}
\appendix

\section{MLS-MPM and constitutive model}
\label{apd:a}

We analyze the foam compression problem (Sec.~\ref{sec:MPM}) using MLS-MPM~\cite{hu2018moving}, which provides improved accuracy and stability through moving least squares reconstruction. The computational domain is discretized on a background grid with resolution $n_{\text{grid}} \times n_{\text{grid}}$ and grid spacing $\Delta x = 1/n_{\text{grid}}$ in normalized coordinates. Material particles are initialized uniformly within grid cells with $n_{\text{ppc}}$ particles per cell dimension, yielding particle spacing $\Delta x / n_{\text{ppc}}$ and area $A_{p,0} = (\Delta x / n_{\text{ppc}})^2$. Quadratic B-spline basis functions are used for particle-grid transfers to ensure \(C^1\) continuity and reduce numerical dissipation. Each time step involves a particle-to-grid (P2G) transfer to compute grid forces, an explicit grid velocity update, and a grid-to-particle (G2P) transfer using affine particle-in-cell (APIC) interpolation~\cite{jiang2015affine} to update particle velocities, positions, and deformation gradients. The deformation gradient is updated multiplicatively as:
\begin{equation}
\mathbf{F}_p^{n+1} = (\mathbf{I} + \Delta t \mathbf{C}_p^{n+1}) \mathbf{F}_p^n
\end{equation}
where \(\mathbf{C}_p\) is the affine velocity field matrix obtained from grid velocities, which captures the local velocity gradient around each particle. 

The material behavior is described by a finite-strain viscoelastic constitutive model composed of an
equilibrium (hyperelastic) Neo-Hookean response and a non-equilibrium viscous overstress described by
a generalized Maxwell model (Prony series)~\cite{simo1987fully, abaqus2011abaqus}.
Let \(\mathbf F\) denote the deformation gradient and \(J=\det(\mathbf F)\). The equilibrium strain-energy density is
\begin{equation}
\Psi_\infty(\mathbf F)
= \frac{G_\infty}{2}\left(\mathrm{tr}(\mathbf F^{T}\mathbf F)-3-2\ln J\right)
+ \frac{K_\infty}{2}\left(\ln J\right)^2
\end{equation}
where \(G_\infty\) and \(K_\infty\) are the equilibrium shear and bulk moduli, respectively. They are related to the
equilibrium Young's modulus \(E_\infty\) and Poisson's ratio \(\nu_\infty\) by: 
\begin{equation}
G_\infty = \frac{E_\infty}{2(1+\nu_\infty)},\qquad
K_\infty = \frac{E_\infty}{3(1-2\nu_\infty)}
\end{equation}
The corresponding equilibrium first Piola-Kirchhoff stress is:
\begin{equation}
\mathbf P_\infty
= \frac{\partial \Psi_\infty}{\partial \mathbf F}
= G_\infty\left(\mathbf F-\mathbf F^{-T}\right)+K_\infty \ln(J)\,\mathbf F^{-T}
\end{equation}

For the viscous response, the time-dependent shear and bulk relaxation moduli are represented by Prony series:
\begin{equation}
G_R(t)=G_\infty+\sum_{i=1}^{N_g}G_i e^{-t/\tau_i^g},\qquad
K_R(t)=K_\infty+\sum_{j=1}^{N_k}K_j e^{-t/\tau_j^k}
\end{equation}
where \(G_i\) and \(K_j\) are relaxation strengths and \(\tau_i^g\) and \(\tau_j^k\) are the corresponding relaxation times.
The instantaneous moduli are \(G_0=G_\infty+\sum_{i=1}^{N_g}G_i\) and \(K_0=K_\infty+\sum_{j=1}^{N_k}K_j\), and as
\(t\to\infty\), the moduli relax to \(G_\infty\) and \(K_\infty\).

The non-equilibrium (viscous) response is described by internal variables representing the deviatoric Cauchy overstresses \(\mathbf q_i\) and volumetric Cauchy overstresses \(q_j\):
\begin{equation}
\mathbf P_{\mathrm{vis}} = J (\sum_{i=1}^{N_g}\mathbf q_i+\sum_{j=1}^{N_k} q_j\,\mathbf I)\mathbf F^{-T}
\end{equation}

The internal variables evolve according to first-order kinetics:
\begin{equation}
\overset{\triangle}{\mathbf q}_i+\frac{1}{\tau_i^g}\mathbf q_i = 2G_i\,\mathbf D_{\mathrm{dev}},
\qquad
\dot q_j+\frac{1}{\tau_j^k} q_j = K_j\,\mathrm{tr}(\mathbf D)
\end{equation}
where \(\mathbf D=\tfrac12(\mathbf L+\mathbf L^{T})\) is the rate-of-deformation tensor and \(\mathbf D_{\mathrm{dev}}=\mathbf D-\tfrac13\mathrm{tr}(\mathbf D)\mathbf I\) is the deviatoric part. \(\mathbf W=\tfrac12(\mathbf L-\mathbf L^{T})\) is the spin tensor, where \(\mathbf L=\dot{\mathbf F}\mathbf F^{-1}\) is the velocity gradient. \(\overset{\triangle}{\mathbf q}_i\) is the Jaumann objective rate, given by $\overset{\triangle}{\mathbf q}_i = \dot{\mathbf q}_i + \mathbf q_i\mathbf W-\mathbf W\mathbf q_i$.

Finally, the total first Piola-Kirchhoff stress is obtained as:
\begin{equation}
\mathbf P = \mathbf P_\infty + \mathbf P_{\mathrm{vis}}
\end{equation}

For the 2D plane strain formulation, each particle \(p\) contributes a force to grid node \(i\):
\begin{equation}
\mathbf{f}_i = -\sum_p A_{p,0} \mathbf{P}_p \nabla w_{ip}
\end{equation}
where \(A_{p,0}\) is the initial area of particle \(p\) in the reference configuration, \(\mathbf{P}_p\) is its first Piola-Kirchhoff stress tensor, and \(\nabla w_{ip}\) is the gradient of the quadratic B-spline weight function evaluated at particle \(p\) with respect to node \(i\). 

The material properties of the photosensitive resin and the parameters of the MLS-MPM model used in Sec.~\ref{sec:MPM} are listed in Table.~\ref{Tab:params}.

\begin{table}[H]
\centering
\caption{Material and simulation parameters used in the MLS-MPM compression simulation.}
\begin{tabular}{lll}
\toprule
\text{Property and parameter} & \text{Value} & \text{Unit} \\
\midrule
Density ($\rho$) & $1150$ & kg$\cdot$m$^{-3}$ \\
Young's modulus ($E$) & $1.8 \times 10^{6}$ & Pa \\
Poisson's ratio ($\nu$) & $0.4$ &  \\
Gravity ($\mathbf{g}$) & $\mathbf{0}$ & m$\cdot$ $s^{-2}$ \\
Shear relaxation ($[G_i, \tau_i^g]$) & [0.25$G_\infty$, 0.5] & [Pa, s] \\
Bulk relaxation ($[K_i, \tau_i^k]$) & [0.0, 0.0] & [Pa, s] \\
Compression velocity ($v_\text{comp}$) & 0.005 & m$\cdot s^{-1}$ \\
Time step ($\Delta t$) & $3 \times 10^{-4}$ & s \\
Foam dimension & 30 $\times$ 30 & mm$^2$ \\
Grid resolution ($n_\text{grid}$) & 104 &  \\
Particle density ($n_\text{ppc}$) & 2 &  \\
Padding layer & 20 &  \\
\bottomrule
\end{tabular}
\label{Tab:params}
\end{table}

\begin{figure}[H]
    \centering
    \includegraphics[width=0.7\linewidth]{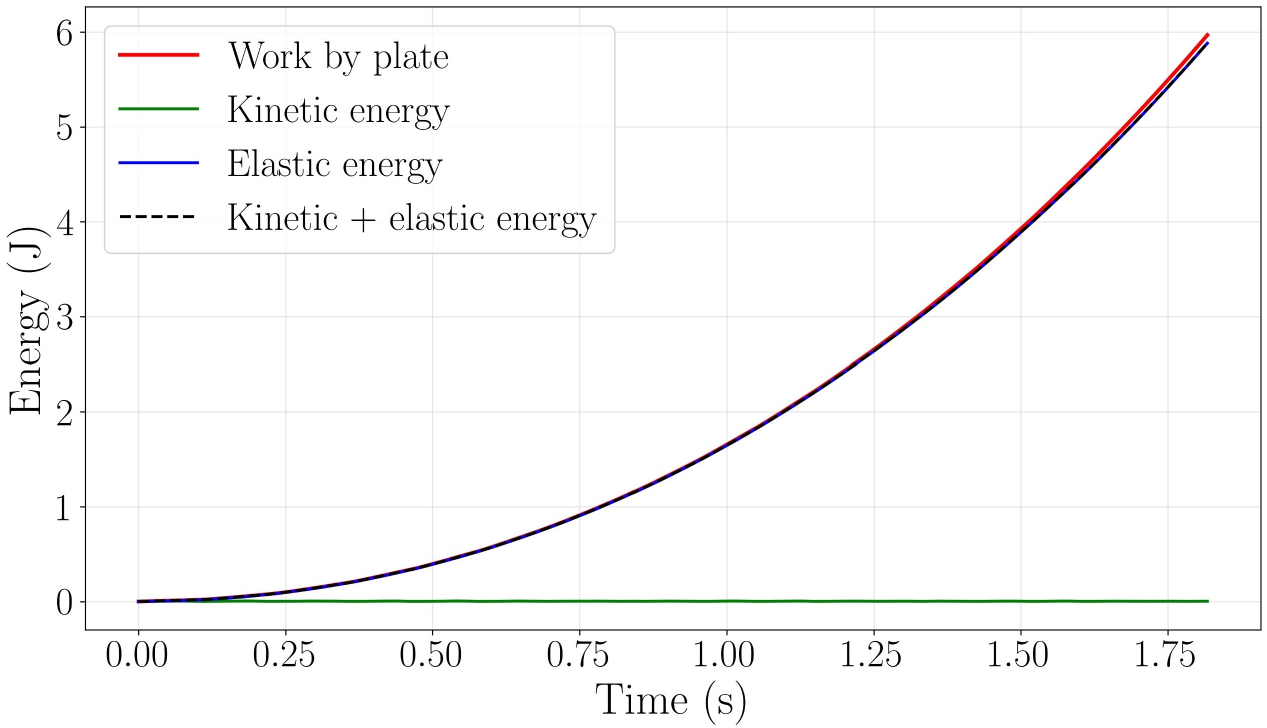}
    \caption{Energy conservation in the compression simulation with MLS-MPM.}
    \label{fig:eng_con}
\end{figure}

To validate the accuracy of the simulation framework, we use only the hyperelastic model to analyze energy conservation during compression of a 2D cubic block made of the photosensitive resin to 30\% strain. The energy balance can be expressed as:
\begin{equation}
    W_\text{plate} = E_\text{elastic} + E_\text{kinetic}
\end{equation}

where $W_\text{plate}$ is the work done by the compression plate, and $E_\text{elastic}$ and $E_\text{kinetic}$ are the elastic and kinetic energy of the block, respectively. As shown in Fig.~\ref{fig:eng_con}, the temporal evolution of these energies demonstrates good conservation, with an energy loss rate of 1.41\% over the entire compression process, primarily attributed to numerical dissipation and time discretization errors.

\section{Frictional self-contact in MPM}
\label{apd:b}

The damage-field gradient (DFG) partitioning method addresses fundamental limitations in the Material Point Method (MPM) regarding localization and frictional contact by dynamically partitioning material into contact pairs based on gradients of a scalar field constructed from particle data~\cite{homel2017field}.

A continuous damage field is constructed from discrete particle damage measures through kernel interpolation. The unnormalized damage field at spatial position $\mathbf{x}$ is defined as:
\begin{equation}
\mathcal{D}(\mathbf{x}) = \sum_{p=1}^{n_p} D_p \omega\left(\frac{|\mathbf{x} - \mathbf{x}_p|}{r_p}\right)
\end{equation}
where $D_p$ is the scalar damage measure at particle $p$, $\mathbf{x}_p$ is the particle position, and $\omega$ is a spherically symmetric cubic kernel with support radius $r_p$:
\begin{equation}
\omega(\bar{r}) = \begin{cases}
1 - 3\bar{r}^2 + 2\bar{r}^3 & 0 \leq \bar{r} \leq 1 \\
0 & \bar{r} > 1
\end{cases}
\end{equation}

To eliminate spurious edge effects inherent in kernel summation methods, the damage field is normalized by the kernel sum field:
\begin{equation}
\bar{\mathcal{D}}(\mathbf{x}) = \frac{\mathcal{D}(\mathbf{x})}{S(\mathbf{x})}, \quad S(\mathbf{x}) = \sum_{p=1}^{n_p} \omega\left(\frac{|\mathbf{x} - \mathbf{x}_p|}{r_p}\right)
\end{equation}

Direct evaluation of the damage field gradient at grid nodes, $\nabla\bar{\mathcal{D}}(\mathbf{x}_i)$, proves inadequate when the node coincides with damage field extrema, yielding $\nabla\bar{\mathcal{D}}(\mathbf{x}_i) = \mathbf{0}$. To circumvent this limitation, a nonlocal gradient operator is defined. Conceptually, the nonlocal gradient direction at node $i$ is determined by identifying the location within neighborhood $\mathcal{R}_i$ where the gradient magnitude is maximal:
\begin{equation}
\mathbf{x}_i^{\text{max}} = \arg\max_{\mathbf{x} \in \mathcal{R}_i} |\nabla \bar{\mathcal{D}}(\mathbf{x})| \quad \text{subject to} \quad |\mathbf{x} - \mathbf{x}_i| < r_g
\end{equation}
where $r_g$ is the search radius, typically taken as the grid cell diagonal length. The nonlocal gradient is then:
\begin{equation}
\nabla \bar{\mathcal{D}}_i^* = \nabla \bar{\mathcal{D}}(\mathbf{x}_i^{\text{max}})
\end{equation}

Material partitioning into distinct velocity fields is performed independently at each grid node based on the relative orientation of particle and nodal damage gradients. For a particle $p$ contributing to node $i$, the velocity field assignment $\zeta \in {0, 1}$ is determined by:
\begin{equation}
\zeta = \begin{cases}
0 & \text{if } \nabla\bar{\mathcal{D}}_p \cdot \nabla\bar{\mathcal{D}}_i^* \geq 0 \\
1 & \text{if } \nabla\bar{\mathcal{D}}_p \cdot \nabla\bar{\mathcal{D}}_i^* < 0
\end{cases}
\end{equation}

The partitioned mass and momentum at node $i$ for velocity field $\zeta$ are then:
\begin{equation}
m_{\zeta i} = \sum_{\zeta p} \bar{S}_{ip}^* m_p, \quad \mathbf{q}_{\zeta i} = \sum_{\zeta p} \bar{S}_{ip}^* m_p \mathbf{v}_p
\end{equation}
where $\sum_{\zeta p}$ denotes summation over particles assigned to field $\zeta$ at node $i$, and $\bar{S}_{ip}^*$ is the shape function.

Two separability criteria determine whether material at a node should be treated as a cohesive continuum or as distinct contacting surfaces. Two conditions must be satisfied simultaneously:

\textbf{Condition 1 (Complete failure):} At least one particle mapping to the node must be fully damaged:
\begin{equation}
\max\left(D_{1i}^{\text{max}}, D_{2i}^{\text{max}}\right) = 1
\end{equation}
where $D_{\zeta i}^{\text{max}} = \max_{p \in \zeta} D_p$ for particles in field $\zeta$ at node $i$.

\textbf{Condition 2 (Surface formation):} The minimum mass-weighted average damage across both fields must exceed a threshold:
\begin{equation}
\min(\bar{D}_{1i}, \bar{D}_{2i}) > \bar{D}^{\text{min}}
\end{equation}
where the field-averaged damage is:
\begin{equation}
\bar{D}_{\zeta i} = \frac{\sum_{\zeta p} \bar{S}_{ip}^* D_p m_p}{\sum_{\zeta p} \bar{S}_{ip}^* m_p}
\end{equation}

When the separability criterion is satisfied, standard MPM multi-field contact algorithms are applied.

\begin{figure}[H]
    \centering
    \includegraphics[width=1.0\linewidth]{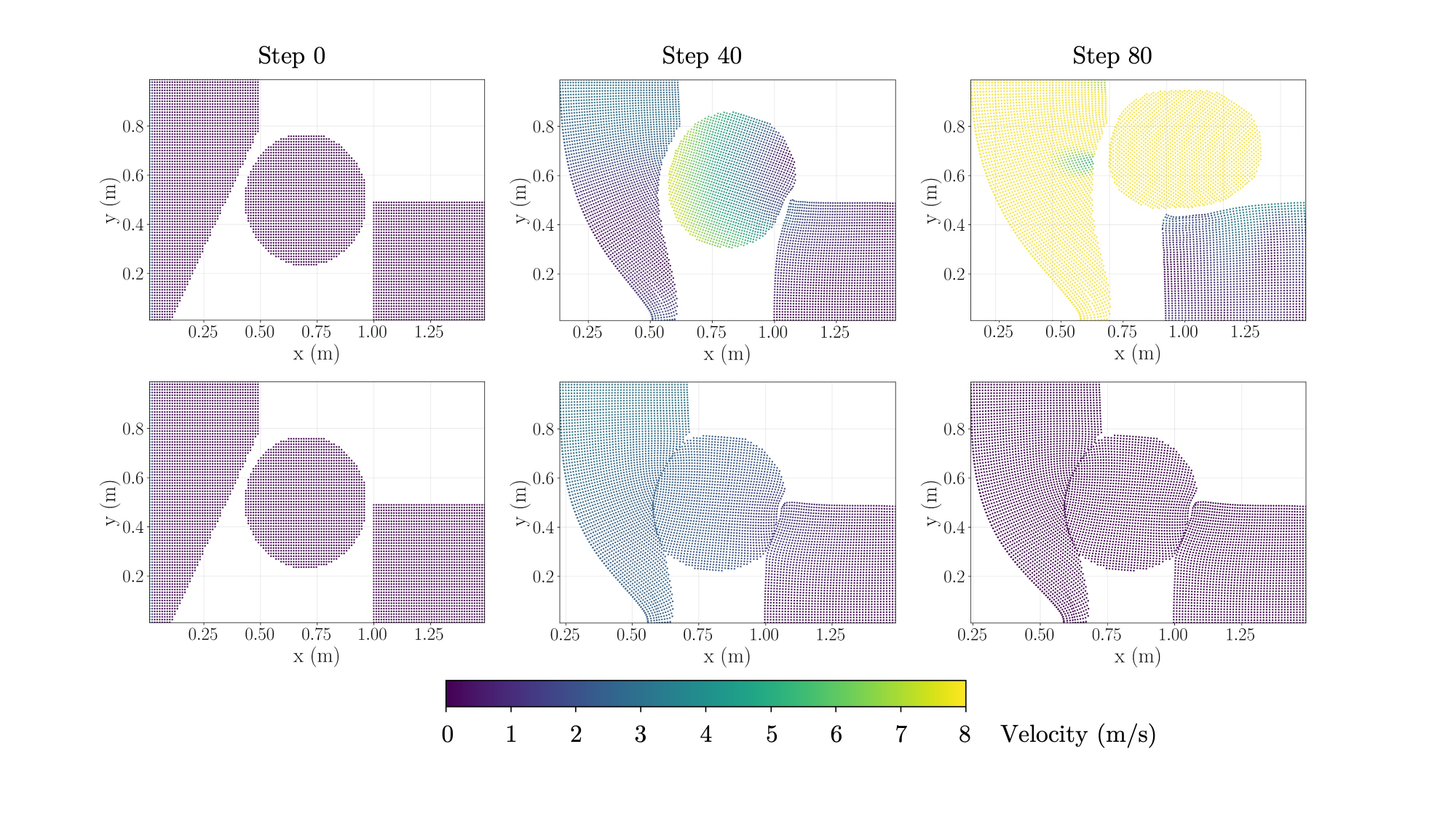}
    \caption{Large-deformation plane-strain frictional contact example showing the initial configuration (left) and contours of speed (center, right). Top: field-gradient partitioning self-contact allows relative motion between contact surfaces. Bottom: single-field solution produces no-slip contact.}
    \label{fig:con_bench}
\end{figure}

For self-contact problems such as foam compression, surface particles are assigned $D_p = 1$ while interior particles have $D_p = 0$. This enables automatic handling of complex contact as surfaces deform and come into proximity. A benchmark study demonstrating this approach is presented in Fig.~\ref{fig:con_bench}.

\section{Phase field brittle fracture}
\label{apd:c}
Within the phase field framework, the fracture evolution can be characterized through minimization of the total energy functional with respect to the displacement field $\mathbf{u}$ and the phase field variable $\zeta$:
\begin{align}
    \Pi(\mathbf{u}, \zeta) = \int_{\Omega} \psi(\mathbf{\boldsymbol{\varepsilon}}(\mathbf{u}), \zeta)\, \mathrm{d}\Omega + \int_{\Omega}g_c\gamma(\zeta, \nabla \zeta)\, \mathrm{d}\Omega  - \int_{\Omega}\mathbf{b}\cdot\mathbf{u}\, \mathrm{d}\Omega - \int_{\Gamma_N}\mathbf{t}\cdot\mathbf{u}\, \mathrm{d}\Gamma
\end{align}
where $g_c$ denotes the critical energy release rate in the sense of Griffith and $\gamma$ represents the regularized crack surface density function per unit volume: 
\begin{align}
    \gamma(\zeta, \nabla \zeta) = \frac{1}{2l}\zeta^2 + \frac{l}{2} |\nabla \zeta|^2
\end{align}
where $l$ denotes the length-scale parameter. The elastic strain energy density $\psi$ is expressed as:
\begin{align}
    \psi(\boldsymbol{\varepsilon}, \zeta) = g(\zeta) \psi_{+}(\mathbf{\boldsymbol{\varepsilon}}) + \psi_{-}(\mathbf{\boldsymbol{\varepsilon}})
\end{align}
where $g(\zeta) = (1-\zeta)^2$ is the degradation function which characterizes the progressive loss of material stiffness induced by crack formation. We employ the spectral decomposition approach proposed by Miehe et al.~\cite{miehe2010phase} for the strain tensor:
\begin{align}
    \psi_{\pm}(\mathbf{\boldsymbol{\varepsilon}}) &= \frac{\lambda}{2}\langle\mathrm{tr}(\mathbf{\boldsymbol{\varepsilon}})\rangle^2_{\pm} + \mu\mathbf{\boldsymbol{\varepsilon}}_{\pm}:\mathbf{\boldsymbol{\varepsilon}}_{\pm}
\end{align}
where $\mathbf{\boldsymbol{\varepsilon}}_{\pm}:=\sum_{a=1}^n \langle \varepsilon_a \rangle_{\pm} \mathbf{n}_a \otimes \mathbf{n}_a$, with $\{ \varepsilon_a \}_{a=1}^n$ and $\{ \mathbf{n}_a \}_{a=1}^n$ representing the principal strains and principal directions of $\mathbf{\boldsymbol{\varepsilon}}$, respectively. The Macaulay bracket is defined as $\langle x \rangle_{\pm}:=\frac{1}{2}(x\pm|x|)$. 

The material properties and phase field parameters used for Sec.~\ref{sec:pf} are listed in Table.~\ref{Tab:params2}.

\begin{table}[H]
\centering
\caption{Material and simulation parameters used in the phase field fracture simulation.}
\begin{tabular}{lll}
\toprule
\text{Property and parameter} & \text{Value} & \text{Unit} \\
\midrule
Phase 1 Young's modulus ($E_1$) & $2.1 \times 10^{5}$ & $\mathrm{kN/mm^2}$ \\
Phase 2 Young's modulus ($E_2$) & $2.1 \times 10^{3}$ & $\mathrm{kN/mm^2}$ \\
Poisson's ratio ($\nu$) & $0.3$ &  \\
Energy release rate ($g_c$) & $2.4 \times 10^{-3}$ & kN/mm \\
Length-scale parameter ($l$) & 0.05 & mm \\
Displacement ($u$) & 0.01 & mm \\
Simulation steps & 100 & \\
Foam dimension & 1 $\times$ 1 & mm$^2$ \\
Grid resolution & 64 $\times$ 64 & \\
\bottomrule
\end{tabular}
\label{Tab:params2}
\end{table}




\end{document}